# Permanent magnets including undulators and wigglers

*J. Bahrdt*
Helmholtzzentrum für Materialien und Energie, Berlin, Germany

**Abstract**
After a few historic remarks on magnetic materials we introduce the basic definitions related to permanent magnets. The magnetic properties of the most common materials are reviewed and the production processes are described. Measurement techniques for the characterization of macroscopic and microscopic properties of permanent magnets are presented. Field simulation techniques for permanent magnet devices are discussed. Today, permanent magnets are used in many fields. This article concentrates on the applications of permanent magnets in accelerators starting from dipoles and quadrupoles on to wigglers and undulators.

## 1 History

Permanent magnets were already mentioned around 600 BC by Thales of Miletus. He attributed magnets a soul because he observed that they attract small pieces of iron. People were always impressed by permanent magnet phenomena as documented in an old anecdote: A long time ago a shoemaker walked through the beautiful countryside of Greece. After a while he was surprised that his shoes started to fall apart. Finally, he noticed mysterious stones on his path that pulled out the iron nails from his sandals.

Magnetic materials were discovered two and a half thousand years ago in the Greek area Magnesia, which gave this remarkable material its name — magnetite ($Fe^{II}(Fe^{III})_2O_4$). In 200 BC the first scientific device called Si Nan (translation: pointing south) was built in China. A magnetic 'spoon' spins on a polished surface and once it comes to rest it points southwards. It is not known whether this device ever worked reliably as a compass. The first European compass was mentioned in 1200 AD. A piece of magnetite sitting on a wooden piece in a bowl of water aligns after a while in the north–south direction. Around 1600 William Gilbert described how to magnetize iron by mechanical deformation such as forging or drawing in the north–south direction. Also cooling down a red-hot iron bar may freeze the earth magnetic field. In 1750 the first ferrites were fabricated by Gowan Knight. He used the sintering technique which is still an essential step in today's magnet production. In 1815 Hans Oersted discovered that a current-carrying wire produces a magnet field. Following this invention it took only six years to build the first electromagnet that was capable of magnetizing steel (William Sturgeon). In 1867 a German handbook was published that describes the fabrication of magnetic materials from non-magnetic components and vice versa.

The twentieth century saw a rapid development of various types of magnetic materials which was always driven by the demand for higher remanence, higher stability with respect to reverse fields and temperature, and last but not least by a cost effective production which implied the availability of the constituents of the material. There are only a few chemical elements that show ferromagnetic properties. Those are the transition metals Fe, Co, Ni with Curie temperatures of several 100°C and the lanthanides Eu, Gd, Tb, Dy, Ho, Er, Tm with Curie temperatures below room temperature. All relevant permanent magnets are alloys made of a large variety of components. Over the last 100 years the energy product of permanent magnets increased by several orders of magnitude (Fig. 1).

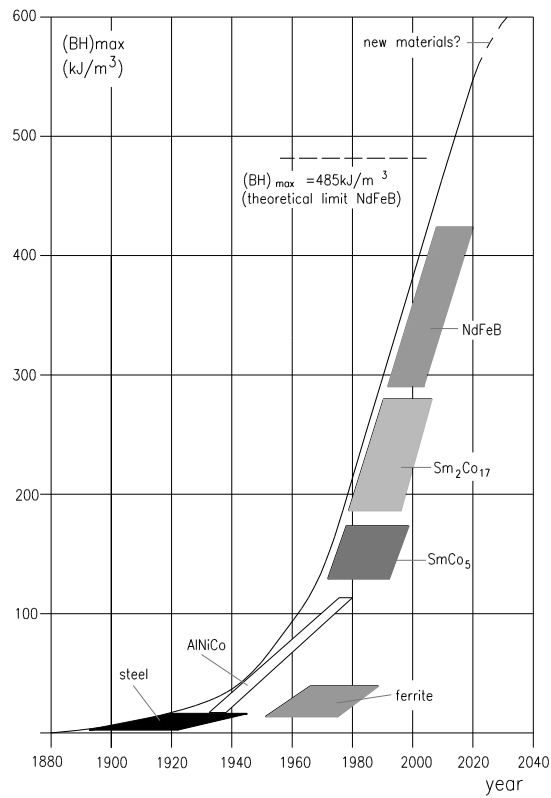

**Fig. 1:** Development of the energy product of various magnet types over the last century [1]

Before coming back in more detail later, we give here a brief overview of the big steps in magnet technology in the twentieth century. In 1916 it was observed that the coercivity of normal steel could be enhanced with Co additions. AlNiCo, an alloy made of Al, Ni and Co, played an important role for several decades. The grades AlNiCo 3 and 5 go back to 1931 and 1938, respectively. In 1938 the production techniques of ferrites were significantly improved in Japan. From 1945 onwards permanent magnets became comparable with electromagnets in performance and cost. New AlNiCo grades with improved performance, AlNiCo 8 and 9, were invented in 1956. Until 1970 AlNiCo was the prevalent permanent magnet material. In 1970 the first rare earth alloy, $SmCo_5$, was produced. Though rare earths are generally more abundant than copper or lead they are not concentrated in big mines and it is difficult to extract the rare earths from the ore. Most of the material is located in China and was not always accessible. Therefore, the research on magnets without rare earths was still ongoing, resulting in the production of FeCrCo in 1971. Owing to the Co crisis in the late 1970s the magnet suppliers started to look for alternatives with less Co content. Hard ferrites were produced all the time since the constituents are plentiful and non strategic. In 1981 $Sm_2(Co,Cu,Fe,Zr)_{17}$ was invented. Pure Sm is expensive, and with the elaboration of a Ca-reduction process Sm-oxides could be used as well, thus reducing the cost significantly. With the invention of $Nd_2Fe_{14}B$ in 1983 a high performing material became available which does not require Co at all. Fe and B are plentiful and Nd is a factor of ten more frequent than Sm. In the following years the $Nd_2FeB_{14}$ production increased exponentially (Fig. 2) and the price per kilogram dropped one order of magnitude (Fig. 3). It is worth mentioning that the USA stopped production in 2004. In Europe only one supplier is left. In 2001 China surpassed Japan in the production rate though most of the material is still used in China. Comparing the tonnage dedicated for export China and Japan are comparable. A trend is observable that magnet suppliers in industrialized countries specialize in downstream products with a higher added value or in high-performance magnets for special applications.

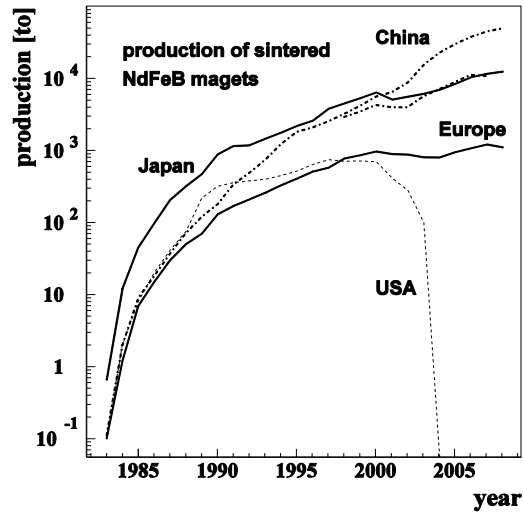

**Fig. 2:** Production rate of $Nd_2Fe_{14}B$ [2]. For China the total magnet production and the production dedicated for export is plotted

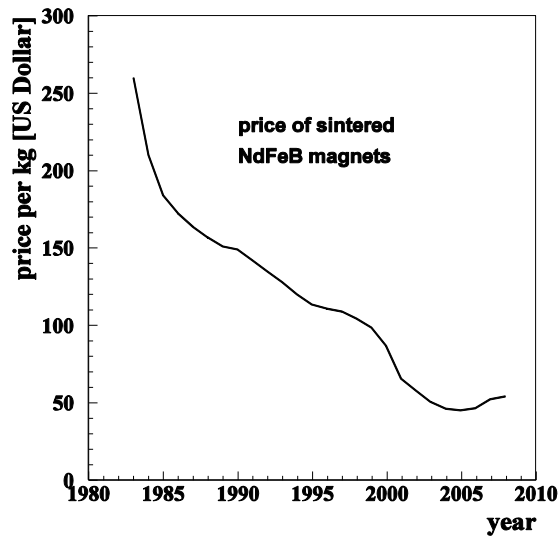

**Fig. 3:** Cost development of $Nd_2Fe_{14}B$ magnets [2]

## 2   Permanent magnet applications

Today, permanent magnets are used in many areas. A systematic classification of the applications related to the underlying physical law is given by R. Parker (see bibliography).

– *Coulomb law:* compass, magnetic bearing, magnetic coupling, fixing tool for machining, transportation line, conveyer, hysteresis device, small MRI system for medical applications.
– *Faraday law:* dynamo, generators based on wind or water energy, microphone, eddy current speedometer.

- *Lorentz force law:* loud-speaker, servo motor, voice coil motor (hard disk drive), device where the Lorentz force acts on free electrons such as: sputter facility, ion getter pump, accelerator magnet including undulator and wiggler, Halbach type dipole and higher order multipole.

The third class of applications will be discussed later in more detail. Figure 4 shows the development of applications from 1999 to 2003. Today, industrialized countries use about 50% of the magnets in voice coil motors. Table 1 gives an overview of permanent magnet applications in China.

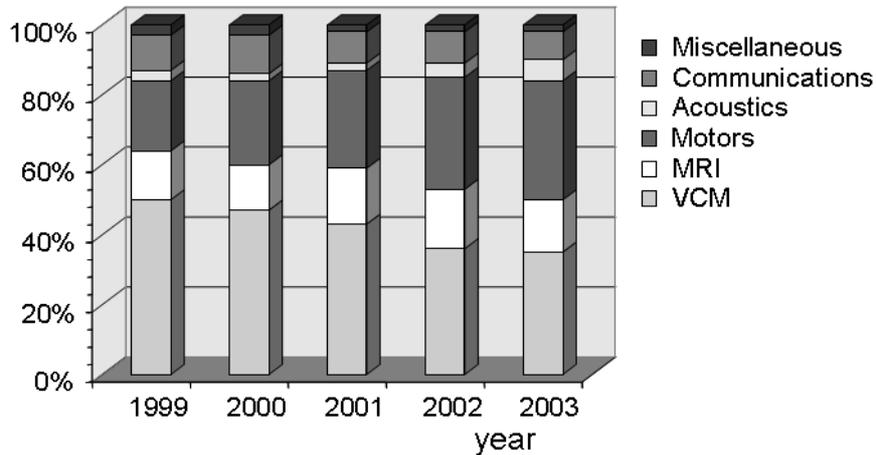

**Fig. 4:** Applications of permanent magnets worldwide in 2004 [3]

**Table 1:** Magnet applications in China in 2007

| High-tech products (t) | | Low-tech products (t) | |
|---|---|---|---|
| MRI | 1800 | Loud speaker | 11280 |
| VCM | 1300 | Separator | 3610 |
| CD-pickup | 2515 | Magnetizer | 900 |
| DVD / CD-ROM | 4060 | | |
| Mobile phone | 3160 | | |
| Cordless tool | 3180 | | |
| Electric bike | 5860 | | |

There are several advantages of permanent magnets as compared to electromagnets: i) Machines such as motors can be built more compactly with permanent magnets. A permanent magnet array can be scaled in all three dimensions maintaining the magnetic field level at the centre. This is possible since permanent magnets can be described as blocks that carry infinite thin layers of surface currents with a constant surface current density of the order of 10 kA/cm (see Section 10). Scaling an electromagnet to smaller dimensions while keeping the field constant requires a linear enhancement of the current density. The technical limit for water-cooled coils is about 500 A/cm$^2$. ii) In principle, infinitely high fields can be produced. Let us imagine a magnet configuration that produces a field $B_0$. Scaling this configuration by a factor of two in all three dimensions and adding it to the old configuration doubles the total field. Of course, the available space and price limit this procedure rather soon. Nevertheless, there are many devices whose fields exceed the remanent field by large factors, e.g., Iwahita built a 3.9 T dipole magnet [4]. iii) The power consumption is zero and, thus, there are no cooling problems. iv) Permanent magnet devices are failsafe as long as they are operated below the Curie temperature.

## 3 Basic definitions

In this article we use Gaussian units. The macroscopic property of magnetic material is described by the dependence of magnetic induction and magnetization on the external field. The zero crossing of the induction or magnetization between the second and third quadrant is called coercive force (the absolute value of the external field as plotted in Fig. 5). The location of this crossing can be significantly different for the induction ($H_c$) and the magnetization ($H_{cj}$). The *BH*-dependency is reproduced if the material is periodically driven to complete magnetization in one direction and complete magnetization in the opposite direction by tuning the external field. A non-magnetized block shows a different dependency which is called the initial magnetization curve. The local dependence between *B* and *H* is described by the permeability. Apart from the usual permeability $\mu = dB/dH$ which describes the linear part in the first and second quadrant we differentiate between the initial, the differential or maximum, and the reversal or recoil permeability (Fig. 6). The reversal permeability describes the properties in the non-linear part close to the knee. For small field variations the *BH*-curve exhibits a small loop around a straight line which is approximately parallel to the slope of the *BH*-curve above the knee. $Nd_2Fe_{14}B$ magnets have a different permeability parallel and perpendicular to the easy axis of about $\mu_{par} = 1.04$ and $\mu_{perp} = 1.17$. Depending on the working point of the magnet, i.e., the operating point on the hysteresis loop, the remanence can be lower than $B_r = B(H = 0)$ by a few per cent.

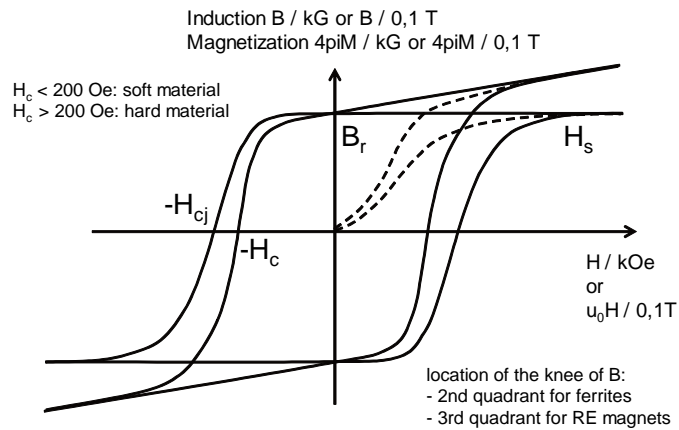

**Fig. 5:** Hysteresis loop of a magnetic material

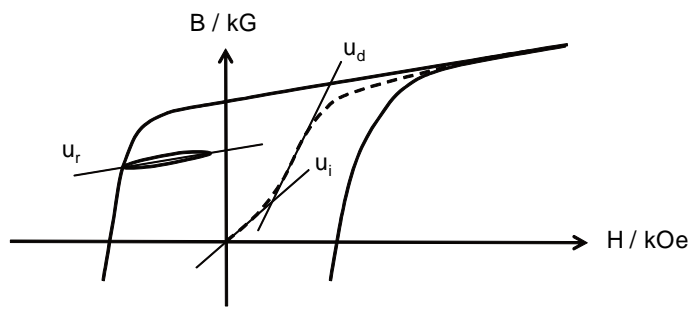

**Fig. 6:** Initial permeability $\mu_i$, differential or maximum permeability $\mu_d$, and reversal or recoil permeability $\mu_r$

We call a material hard magnetic if $H_c > 200$ Oe which is a bit arbitrary. The performance of a magnet is related to the maximum energy product. It is equal to the largest rectangle which can be plotted under the $B$H-curve (Fig. 7). Though modern magnets have a nearly rectangular $BH$-curve we have always

$$(BH)_{max} \leq B_r^2 / \mu.$$

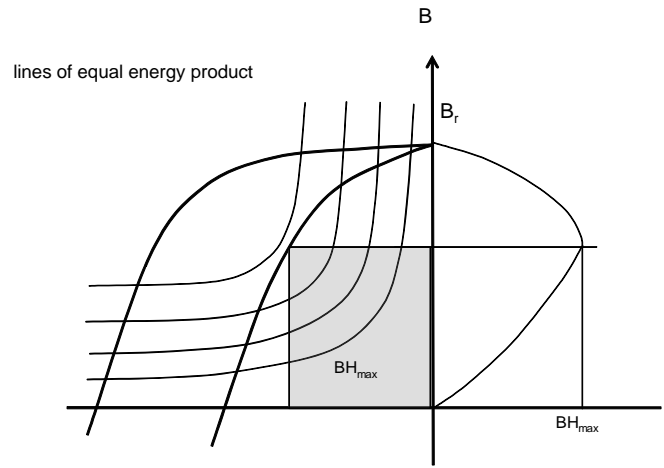

**Fig. 7:** Two graphic representations of the maximum energy product

The efficiency of an electric circuit is given by the conductivity = 1 / resistance = current / voltage. In analogy the efficiency of a magnetic circuit is written as: permeance = 1 / reluctance = flux / magnetomotive force difference. The magnetomotive force is the magnetic potential as produced by currents or magnetized samples. The permeance of a volume $V$ is defined by

$$P = \frac{\iint \vec{B} \cdot d\vec{s}}{\int \vec{H} \cdot d\vec{l}}$$

where the integral over $B$ is taken at the plane $B$ and the line integral over $H$ is taken between the planes A and C (Fig. 8)

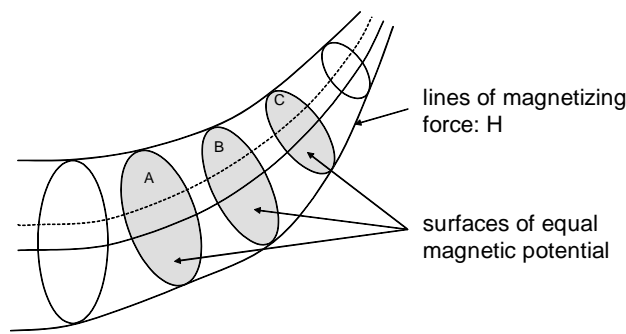

**Fig. 8:** For the definition of the permeance see text

Inside a magnetized body demagnetizing forces proportional to the magnetization are present. The strength is given by the demagnetization factor $D$ which varies between 0 and $4\pi$ depending on the shape of the sample (small for pencil-like samples with the easy axis pointing in the long direction and large for flat samples). The working point of a material is given by $B_d$ and $H_d$ where $B_d / H_d$ is the coefficient of demagnetization or unit permeance. The line connecting the origin and the working point is called the load line. In a magnetic design the working point has to be chosen such that it is well above the knee for all possible operation conditions. Knowing the demagnetization factor the working point can be derived according to:

$$\vec{B} = \vec{H} + 4\pi \cdot \vec{M}$$

$$\vec{H}_d = -D \cdot \vec{M}$$

$$\vec{B}_d = \vec{H}_d - \frac{4\pi}{D}\vec{H}_d$$

$$\frac{\vec{B}_d}{\vec{H}_d} = 1 - \frac{4\pi}{D}$$

In Cartesian coordinates we always have: $D_x + D_y + D_z = 4\pi$. For specific geometries the literature provides tables and approximated expressions for the evaluation of demagnetization factors $D$ (see the book by R. Parker in the bibliography). More complicated geometries have to be evaluated numerically or measured. Depending on the measurement method one gets either magnetometric or fluxmetric (ballistic) demagnetization factors. In the first case a long coil is placed along the complete sample yielding averaged values for the whole sample, in the second case a flat coil is placed around the centre cross section yielding a value averaged over this area. The magnetometric values are always larger than the fluxmetric values.

Analytic expressions of demagnetization factors are available for a generalized ellipsoid and for a rectangular prism. Osborn [5] gives values for a generalized ellipsoid with the semi-axes $a \geq b \geq c$:

$$D_x = \frac{4\pi \cos(\varphi)\cos(\vartheta)}{\sin^3(\vartheta)\sin^2(\alpha)}(F(k,\vartheta) - E(k,\vartheta))$$

$$D_y = \frac{4\pi \cos(\varphi)\cos(\vartheta)}{\sin^3(\vartheta)\sin^2(\alpha)\cos^2(\alpha)}\left(E(k,\vartheta) - F(k,\vartheta)\cos^2(\alpha) - \frac{\sin^2(\alpha)\sin(\vartheta)\cos(\vartheta)}{\cos(\varphi)}\right)$$

$$D_z = \frac{4\pi \cos(\varphi)\cos(\vartheta)}{\sin^3(\vartheta)\cos^2(\alpha)}\left(\frac{\sin(\vartheta)\cos(\varphi)}{\cos(\vartheta)} - E(k,\vartheta)\right)$$

$$\cos(\vartheta) = c/a$$

$$\cos(\varphi) = b/a$$

$$\sin(\alpha) = \sin(\varphi)/\sin(\vartheta) = k$$

$F$ and $E$ are elliptical integrals of the first and second kind with $k$ = modulus and $\theta$ = amplitude. It has to be emphasized that these values are exact and they are constant over the whole volume. Special cases are the sphere with $D_x = D_y = D_z = 4\pi/3$, the infinite long circular cylinder with $D_{par} = 0$ and $D_{perp} = 2\pi$ or an infinite wide plane with $D_{in\text{-}plane} = 0$ and $D_{perp\text{-}plane} = 4\pi$.

Averaged demagnetization factors for a parallel prism are derived by Aharoni [6]. In reality the factors vary over the volume.

$$D_z/4 = \frac{b^2-c^2}{2bc}\ln\left(\frac{sabc-a}{sabc+a}\right) + \frac{a^2-c^2}{2ac}\ln\left(\frac{sabc-b}{sabc+b}\right) + \frac{b}{2c}\ln\left(\frac{sab+a}{sab-a}\right) + \frac{a}{2c}\ln\left(\frac{sab+b}{sab-b}\right)$$

$$+ \frac{c}{2a}\ln\left(\frac{sbc-b}{sbc+b}\right) + \frac{c}{2b}\ln\left(\frac{sac-a}{sac+a}\right) + 2\arctan\left(\frac{ab}{c \cdot sabc}\right) + \frac{a^3+b^3-2c^3}{3abc}$$

$$+ \frac{a^2+b^2-2c^2}{3abc}sabc + \frac{c}{ab}(sac+sbc) - \frac{sab^3+sbc^3+sac^3}{3abc}$$

$$sabc = \sqrt{a^2+b^2+c^2}$$
$$sab = \sqrt{a^2+b^2}$$
$$sac = \sqrt{a^2+c^2}$$
$$sbc = \sqrt{b^2+c^2}$$

In analogy $D_x$ and $D_y$ can be derived. Special cases are a cube with $D_x = D_y = D_z = 4\pi/3$ and an infinite long rectangular cylinder with

$$D_{par} = 0$$

$$D_{perp}/4 = \frac{1-p^2}{2p}\ln(1+p^2) + p \cdot \ln(p) + 2 \cdot \arctan(1/p)$$

$$p = c/a$$

For more complicated geometries or if the demagnetization distribution over a rectangular block is needed, analytic expressions do not exist and the demagnetization factors have to be evaluated numerically. Figure 9 shows the variation of the demagnetization over rectangular blocks with different shapes.

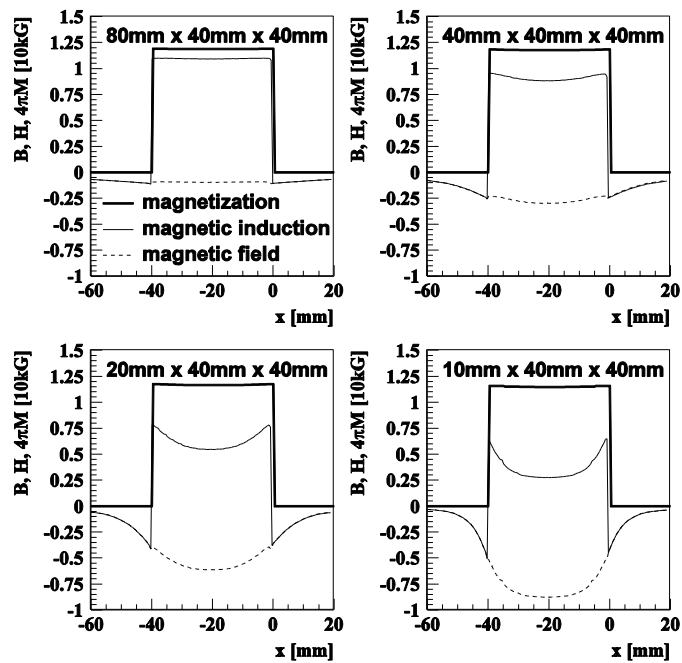

**Fig. 9:** Magnetic induction, magnetic field, and magnetization for four blocks with different geometries. The length in the easy axis direction is varied as indicated in the graph

# 4 Permanent magnet types

In this section the magnetic properties of the most common magnet types are summarized. This overview will give only a taste of the diversity of magnet properties. The magnet material has to be chosen carefully according to the demands: Some applications need high remanence, others require high coercivity or high temperature stability. Some applications do not require high performance and only the price counts. The properties of high-end grades are listed in Tables 2–9.

Permanent magnets are either of type I or type II (Fig. 10). Type I magnets have a high leakage flux leaving the magnet at the sides. The energy stored in these leakage fields is not usable. The permeability of these materials is large and $H_{cj}$ is usually smaller than $B_r$. Typical examples are 35%Co Fe or AlNiCo. Type II magnets have a low leakage flux, the permeability is close to one and $H_{cj}$ is much larger than $B_r$. Rare earths based magnets as well as hard ferrites belong to this species. Sintered magnets are either isostatic pressed (IP), transversally pressed (TP), or axially pressed (AP).

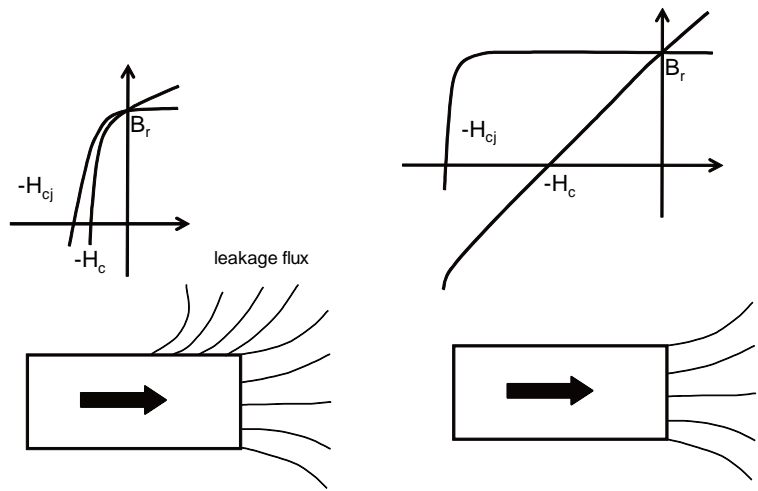

**Fig. 10:** Type I magnets (left) and type II magnets (right)

## 4.1 Carbon steel or martensitic steel

Pure carbon steel (up to one per cent carbon) has a rather low energy product which can be significantly enhanced with the addition of Co (Fig. 11). The performance can further be improved with other non-magnetic ingredients. Internal strain and lattice imperfections can also have a positive effect on the performance.

**Table 2:** Magnetic properties of carbon steel

| Grade | Remanence (kG) | Coercivity $H_{cj}$ (kOe) | Energy product (MGOe) |
|---|---|---|---|
| 3.5 Cr | 9.8 | 0.05 | 0.22 |
| 36.0 Co | 9.6 | 0.24 | 0.94 |

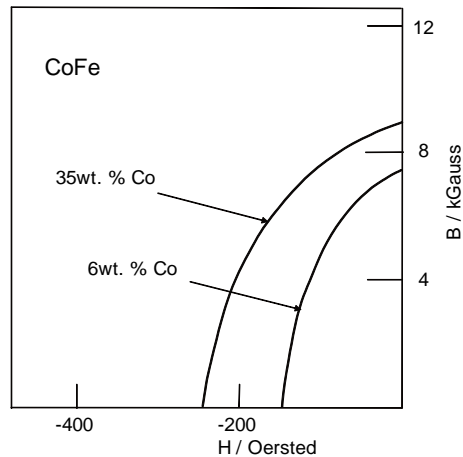

**Fig. 11:** Carbon steel

### 4.2 AlNiCo

The magnets consist of an alloy of Al, Ni, Co and also Fe, Cu, Ti. The remanence is pretty high whereas the coercivity is low which has to be taken into account in the magnetic design. Some grades can be operated at temperatures up to 550°C. The energy product can be enhanced when an anisotropy is deliberately introduced, e.g., by cooling the blocks in a magnetic field. The material is extremely difficult to machine and very brittle which requires a near to finish production. Better mechanical properties have been achieved with sintered magnets. The temperature coefficients are as low as -0.02% for $B_r$ and -0.02% up to 0.01% for $H_{cj}$.

**Table 3:** Typical magnetic properties of a few AlNiCo grades

| Grade | Remanence (kG) | Coercivity $H_{cj}$ (kOe) | Energy product (MGOe) |
| --- | --- | --- | --- |
| AlNiCo 5 cast | 13.5 | 0.74 | 7.5 0 |
| AlNiCo 9 cast | 10.6 | 1.5 | 10.0 |
| AlNiCo 5 sintered | 11.2 | 0.61 | 4.4 |
| AlNiCo 8 sintered | 8.0 | 1.60 | 4.5 |

### 4.3 FeCoCr

The material properties are similar to those of AlNiCo 5 but this magnet requires less Co. Furthermore, it has a higher ductility and it can be oriented by mechanical deformation.

**Table 4:** Magnetic properties of FeCoCr

| Grade | Remanence (kG) | Coercivity $H_{cj}$ (kOe) | Energy product (MGOe) |
| --- | --- | --- | --- |
| FeCoCr | 13.0 | 0.55 | 5.0 |

### 4.4 MnCAl

This material does not need Co at all. It has a higher ductility than AlNiCo. The magnet can be oriented in a warm extrusion process. The extrusion is, however, rather expensive.

**Table 5:** Magnetic properties of MnCAl

| Grade | Remanence (kG) | Coercivity $H_{cj}$ (kOe) | Energy product (MGOe) |
|---|---|---|---|
| MnCAl cast | 3.0 | 0.95 | 1.0 |
| MnCAl cast & exruded | 6.0 | 2.5 | 7.0 |

### 4.5 Hard ferrites

The chemical composition is MO 6($F_2O_3$) or $MFe_{12}O_{19}$ with M = Ba, Sr or Pb. The magnets are sintered. They are either isotropic or oriented. The remanence is low but the coercivity is pretty high. The material has large temperature coefficients of -0.2% for the remanence and +0.1 up to 0.5% for the coercivity. $H_c$ and $H_{cj}$ are similar and the knee is located in the second quadrant.

**Table 6:** Magnetic properties of hard ferrites

| Grade | Remanence (kG) | Coercivity (kOe) | Energy product (MGOe) |
|---|---|---|---|
| Hard ferrite | 4.0 | 3.6 (4.0), $H_c$ ($H_{cj}$) | 4.0 |

### 4.6 SmCo$_5$

Samarium based rare earth magnets have a high energy product. The high coercive grade is the favourite material in the presence of strong reverse fields or in an ionizing radiation environment. The maximum operation temperature is 250°C. The temperature coefficients are as low as -0.045% / deg. and -0.22% / deg. ($H_{cj}$).

**Table 7:** Magnetic properties of SmCo$_5$

| Grade | Typ. remanence (kG) | Min. coercivity $H_{cj}$ (kOe) | Typ. energy product (MGOe) |
|---|---|---|---|
| IP | 10.1 | 12.5 | 25.0 |
| IP | 9.0 | 25.0 | 20.0 |

### 4.7 Sm$_2$Co$_{17}$

The material has a higher remanence than SmCo$_5$. The maximum operation temperature is 350–500°C depending on the grade. The temperature coefficients are lower as compared to SmCo$_5$: -0.035% / deg. and -0.12% / deg. ($H_{cj}$).

In comparison with Nd$_2$Fe$_{14}$B magnets, SmCo$_5$ and Sm$_2$Co$_{17}$ magnets are brittle and the handling is delicate. The temperature coefficients are lower as compared to Nd$_2$Fe$_{14}$B which is advantageous in temperature sensitive applications.

Table 8: Magnetic properties of $Sm_2Co_{17}$

| Grade | Typ. remanence (kG) | Min. coercivity $H_{cj}$ (kOe) | Typ. energy product (MGOe) |
|---|---|---|---|
| IP | 11.0 | 20.0 | 28.0 |
| TP | 10.7 | 20.0 | 27.0 |
| AP | 10.4 | 20.0 | 25.0 |

## 4.8 $Nd_2Fe_{14}B$

This material has the highest remanence among all rare earth based permanent magnets. The temperature gradients are higher than in Sm based magnets: -0.09 to -0.11% / deg for $B_r$ and -0.45 to -0.6% for $H_{cj}$. Depending on the amount of added dysprosium the coercivity can be enhanced significantly sacrificing remanence. The maximum operation temperature depends on the grade and varies between 80°C for material of highest remanence and 230°C for material with highest coercivity.

Table 9: Magnetic properties of $Nd_2Fe_{14}B$

| Grade | Typ. remanence (kG) | Min. coercivity $H_{cj}$ (kOe) | Typ. energy product (MGOe) |
|---|---|---|---|
| IP | 14.7 | 11 | 53 |
| TP | 14.1 | 14 | 48 |
| AP | 13.4 | 14 | 43 |

# 5 Temperature dependence

## 5.1 Compensation of temperature dependent effects

Critical permanent magnet applications need a compensation of the temperature dependent remanence changes. This can be accomplished with various strategies:

– Curie alloys consisting of Ni and Fe have a negative temperature gradient of the permeability. They can be used as temperature sensitive flux shunts. With rising temperature they shortcut less flux, and thus compensate for the remanence loss of the permanent magnets. These materials are used, for example, in speedometers. Other applications are accelerator magnets. In the 344 permanent magnet gradient dipoles of the 8.9 GeV antiproton recycler ring at Fermilab, thin Ni-Fe bars shortcut part of the magnetic material (strontium ferrite) between the Fe pole tips and the iron yokes [7].

– The next Brazilian light source, LNLS II, will be based on 48 permanent magnet gradient dipoles made of hard ferrites [8]. The gap of the magnets is passively readjusted with temperature.

– Compared to $Nd_2Fe_{14}B$ material $SmCo_5$ has a small remanence temperature coefficient. It can further be reduced by mixing the material with other compounds with a positive temperature gradient such as $ErCo_5$ and/or $GdCo_5$.

## 5.2 Curie temperature

The temperature range for a safe operation depends on the Curie temperature. Above the Curie temperature the remanence and coercivity drop to zero and the material becomes paramagnetic. The Curie temperatures of some magnetic materials are summarized in Table 10.

**Table 10:** Curie temperatures of typical elements and permanent magnets

| Grade | Curie temperature (°C) |
|---|---|
| Iron | 770 |
| Cobalt | 1130 |
| Ni | 358 |
| $Nd_2Fe_{14}B$ | 310 |
| $SmCo_5$, $Sm_2Co_{17}$ | 700–800 |
| 35% Co steel | 890 |
| CrFeCo | 630 |
| AlNiCo | 850 |
| Hard ferrites | 400 |

The coercivity of $Nd_2Fe_{17}B$ magnets can be enhanced with the addition of Dy (Fig. 12). On the other hand, the Dy reduces the remanence. The maximum operation temperature of $Sm_2Co_{17}$ grades can be raised with a specific tempering procedure. In both cases the Curie temperature of the material grows a bit but much less than the coercivity. Optimized permanent magnets can be used up to about 75% of the Curie temperature.

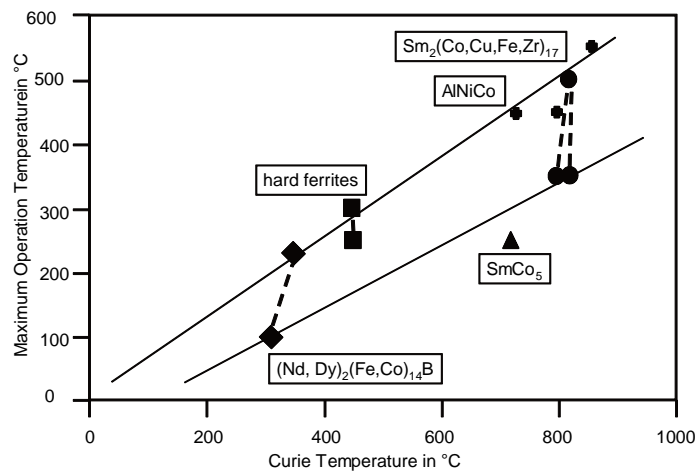

**Fig. 12:** Temperature stability versus Curie temperature of various magnet materials [9]. The dotted lines connect different grades of the same magnet material (for details see text)

Only recently, Hitachi described a vapour deposition and diffusion process which enhances the coercivity of $Nd_2Fe_{14}B$ magnet by about 4 kOe without sacrificing remanence. Alternatively, the remanence can be enhanced by 400 G without losing in coercivity [10]. The magnets are exposed to dysprosium vapour. The Dy atoms diffuse along the grain boundaries into the bulk material without penetrating into the grains. The penetration depth is only a few millimetres which limits this method to thin magnets. This material is of particular interest for in-vacuum undulators where the magnets are close to a several GeV electron beam and the demagnetization stability is an important issue.

### 5.3 Reversible demagnetization

Operating permanent magnets above the Curie temperature causes a complete demagnetization. The crystal structure remains unchanged, and hence the magnets can be remagnetized recovering full performance. Since the demagnetization factor varies over the magnet volume, certain parts demagnetize earlier than more stable regions. For critical applications the magnet supplier deliberately ages the blocks, in order to avoid a remanence loss over the years. For this purpose the magnets can either be heated well above the final operation temperature or they can be stabilized in a reverse field.

Synchrotron radiation light sources are based on periodic permanent magnet structures, so-called undulators, which are passed by relativistic electrons emitting photons in the VUV to X-ray regime (see Sections 12–14). If the magnets are hit by off-axis electrons they may get demagnetized. Demagnetization has been reported at the ESRF [11], the APS [12–13] (see Fig. 13) and DESY [14]. Though the remanence losses reached 10% the structures could be repaired in these cases by re-magnetizing the blocks. If, however, the ionizing radiation changes the crystal structures, irreversible losses may occur [15].

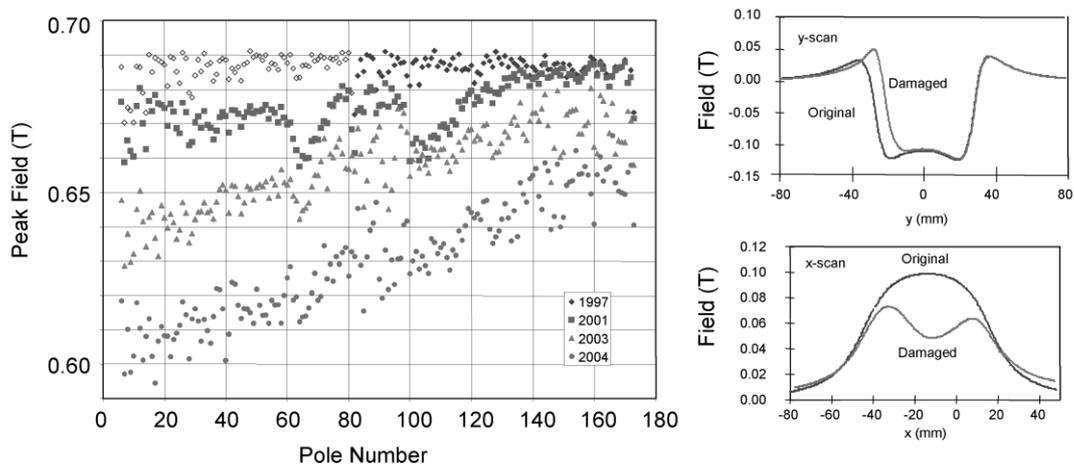

**Fig. 13:** History of sector 3 undulator at the APS (left) and single block Hall probe scans (right). The damage is localized at the surface close to the electron beam (courtesy of L. Moog, APS)

## 6 Permanent magnet fabrication

### 6.1 Sintered magnets

Sumitomo developed the fabrication process for sintered rare earth magnets. The various ingredients are mixed in the desired percentages, melted under vacuum conditions and cast to macroscopic pieces. The pieces are crushed and then milled in several steps down to particle sizes of a few micrometres. The powder is highly reactive and has to be processed under inert gas conditions. The powder is pressed in the presence of a high field which aligns the magnetic domains. The pressed pieces are sintered at temperatures around 850°C which melts the surfaces of the grains (so-called liquid phase sintering). The quality of the magnet depends on the type of pressing.

Highest remanence is achieved with isostatic pressing. Here, the powder is poured into long rubber tubes which are placed into a liquid environment. The powder is compressed with equal pressure from all directions. The magnetic field is applied with a long coil surrounding the tube. These magnets have rather large dipole orientation errors whereas the homogeneity is pretty good. Alternatively, the powder can be pressed in a die. Depending on the geometry, the pressing direction can be parallel (axial pressing) or perpendicular to the field lines (transverse pressing). The latter geometry yields a higher remanence but it can not be applied for block geometries with high aspect ratios. Die pressed magnets have very similar magnetic properties over the whole batch since they all see the same magnetic environment during the pressing. The dipole errors are small but the inhomogeneities can be larger than in isostatic pressed magnets. Die pressing is preferred for large batches because the die can be shaped close to the final geometry and the machining time afterwards can be minimized (near net-shape production). More details on the development and the characteristics of sintered $Nd_2Fe_{14}B$ magnets are given in Ref. [16].

## 6.2 Melt spun magnets

In 1984 General Motors developed another fabrication method for rare earth permanent magnets. The melted ingredients are poured into an induction heated container under Ar atmosphere. A liquid alloy jet is quenched on a water cooled spinning wheel forming a 300 μm microcrystalline ribbon (Fig. 14). This material is further processed following one of three procedures:

– Magnequench I: The ribbons are bonded to form a solid block which can further be machined. The material is isotropic.

– Magnequench II: The ribbons are pressed under high temperature. The material is isotropic.

– Magnequench III: Starting with Magnequench II material the blocks are deformed under high temperature resulting in an anisotropic grade. This material has the highest energy product among the three grades.

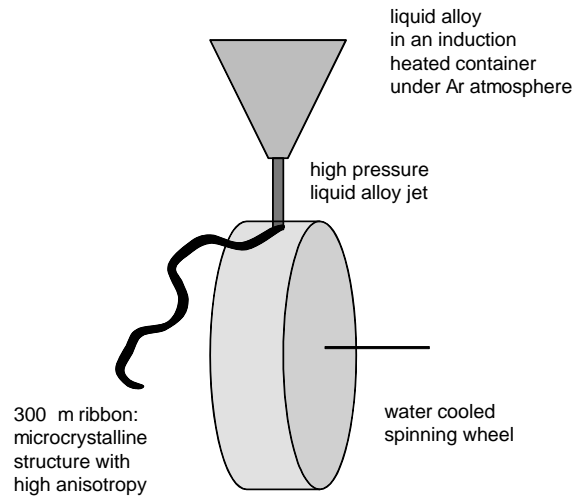

**Fig. 14:** Magnet fabrication following the Magnequench technique

## 7 Measurement techniques for macroscopic properties

The magnetic hysteresis can be measured either in a closed loop or in an open loop geometry. At room temperature or higher temperatures a hysteresis graph can be used (Fig. 15). Here, the sample is clamped between two iron yokes which short-cut the magnetic flux. Powering a coil that is wound around the iron yoke the external field can be varied. The parameters *B* and *H* can be directly measured with pickup coils. Large samples (a few cubic centimetres) can be characterized which minimizes the impact of surface effects.

At low temperatures another method has to be used, i.e., a vibrating sample spectrometer: In a He cryostate a small sample vibrates back and forth inducing a voltage in a pickup coil. The signal is proportional to the dipole moment of the sample. Knowing the sample geometry, this value can be converted to the remanence. Usually, the samples must be small (a few cubic millimetres) and the surface effects have to be taken into account for the interpretation of the results. Furthermore, a precise calibration of the setup is required which includes the demagnetization factor of the sample. Ellipsoidal samples with a constant demagnetization factor are preferable but difficult to fabricate and align.

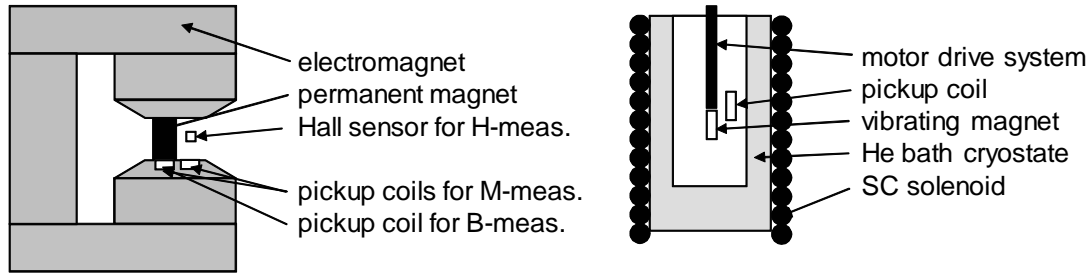

**Fig. 15:** Hysteresis graph (left) and vibrating sample magnetometer (right)

Usually, the dipole moment is measured in a Helmholtz coil with high accuracy. The geometry is optimized such that the measurement is insensitive to a displacement of the magnet and the magnet block size. The magnetizing force of two coils with winding $N$ and current $I$ is

$$H = \frac{2NI}{a}\left[\left(1+\frac{(d/2-x)^2}{a^2}\right)^{-1.5} + \left(1+\frac{(d/2+x^2)}{a^2}\right)^{-1.5}\right]$$

where $d$ = distance of the coils and $a$ = radius of coil.

In the Helmholtz geometry, $d = a$, the quadratic terms disappear. An automated rotation of the magnet around all three axes and an averaging over several turns yields rms reproducibilities better than 0.07% for the main component and better than 0.04° for the easy axis orientation errors with respect to a reference surface (values of the BESSY II system).

The dipole moment can also be measured with a fluxgate sensor located at a large distance (typically 1 m) from the sample. In the far field only the dipole component contributes and higher orders can be neglected. The signals are small and the magnet has to be flipped to get rid of environmental fields. This method is adequate for the main component but less accurate for the minor components.

Close to the magnet surface (i.e., a few millimetres for magnets with dimensions of a few centimetres) the magnet fields can not be described by the dipole moment only. Higher order multipoles have to be included. These higher order terms can be measured with a set-up where the magnet is moved with respect to a fixed wire. The rms reproducibility of the BESSY II system, which uses a single fixed wire, is summerized in Table 11. In other laboratories alternative fluxmetric methods are applied: at SPRING-8 a multifilament rotating coil close to the magnet surface is used [17]. At the ESRF, modules of several magnets are continuously moved at a speed of 60 mm/s passing a fixed multifilament wire (20 single wires) [18]. The modules include an equal number of magnets with the easy axis pointing to the wire and into opposite direction. Thus the modules are magnetically compensated and the field integrals are below 100 G cm. The rms error of a module measurement is 1 G cm.

The information on dipole errors and magnet field inhomogeneities can be used to sort the magnets such that the fabrication errors cancel in the assembled structure.

**Table 11:** Reproducibility of field integral measurements of single magnet blocks (BESSY II system). The blocks have a distance of 5 mm to the wire. The orientation of the easy axis is given with respect to the wire

| Easy axis orientation | Absolute error (rms values) | Relative error (rms values) |
|---|---|---|
| Parallel | $3 \cdot 10^{-4}$ T mm | $6 \cdot 10^{-3}$ |
| Perpendicular | $1.5 \cdot 10^{-3}$ T mm | $2.5 \cdot 10^{-4}$ |

The capacity of the furnace crucible defines the size of a batch of magnets. Within a batch the magnet properties are similar, whereas the dipole moment variations between batches can be as large as a few per cent. Magnets from the same batch can be sorted efficiently resulting in high-performance structures. This strategy is limited to typical batch sizes of 1–2 tonnes. For larger undulators with a total weight of the magnet material of 20 tonnes and more (e.g., undulator length >100 m for FEL structures) other strategies have to be employed. One of them is a sophisticated mixing scheme of the different powder batches before pressing the blocks.

## 8 Microscopic properties

### 8.1 Microscopic structure of RE permanent magnets

Rare earth (RE) based permanent magnets are either sintered or melt spun. They consist of monocrystalline grains with a diameter of a few micrometres which are embedded in a RE-enriched matrix. In the case of $Nd_2Fe_{14}B$ this matrix contains ingredients such as Nd, Co, Cu, Al, Ga, Dy and Nd-oxides. The crystal structures of RE-based magnets are described in detail in Ref. [19]. The unit cell of $SmCo_5$ is hexagonal. $Sm_2Co_{17}$ has a rhombohedral unit cell and $Nd_2Fe_{17}B$ is tetragonal. The monocrystalline areas are formed when the melted phase is cooled down. They remain unchanged during the crushing and milling process and they are oriented with respect to the external field during the pressing. At typical sintering temperatures of about 850°C the matrix enclosing the monocrystalline particles melts whereas the crystallites remain unchanged. This liquid phase sintering yields rather dense products. The theoretical limit of the energy product is given by

$$(BH)_{max} = B_r^2 / \mu$$

$$B_r(20°C) = B_{r-sat}(20°C) \cdot \frac{\rho}{\rho_0} \cdot (1 - V_{nonmagnetic}) \cdot f_\phi ,$$

where

$$f_\phi = \cos(\phi)$$

and

$$\phi = \arctan\left(2 \frac{B_{r-perp}}{B_{r-par}}\right).$$

Typical values for $Nd_2Fe_{14}B$ magnets are as follows [20]. Owing to the liquid phase sintering the density is close to the single crystal density: $\rho/\rho_0 > 0.985$. An optimized pressing process yields alignment coefficients as high as $f_\varphi > 0.98$. The vacuum induction furnace and the consequent inert gas processing keep the amount of impurities (Nd oxides and others) below 2.5 weight per cent. The occurrence of other RE constituents is also below 2.5 weight per cent. Thus the amount of nonmagnetic material is less than 5%. With these parameters an energy product of 59 MGOe has been achieved which is close to the theoretical limit of 63 MGOe.

It is worth noting that $Nd_2Fe_{14}B$ magnets are sensitive to hydrogen decrepitation: $Nd + H_2O \gg NdOH + H$, $H + Nd \gg NdH$. The hydrogen decrepitation can have fatal consequences for magnets operated in a sensitive environment (e.g., undulators of an accelerator). During magnet fabrication special care is needed to cope with these effects. Appropriate chemical constituents between the grain boundaries minimize the hydrogen decrepitation and an appropriate surface passivation or coating (Al, TiN and others) protects the magnets in a humid environment. On the other hand, the decrepitation process may be interesting for a final decomposition of permanent magnet material and a RE recovery [21, 22].

## 8.2  Coercivity

The potential existence of Bloch walls in a magnetic particle depends on the size of the particle. Energy considerations show that Bloch walls can not exist below a certain grain size. For example, this critical size is 0.01 μm for Fe and 1 μm for Ba ferrites. Below these limits the particles behave like a single domain. For larger grains Bloch walls may show up. The typical grain size of a RE-magnet of a few micrometres is a bit larger than the single domain size. Usually, the magnetization vector rotates in the plane of the boundary between two magnetic domains. Exceptions are thin films where the rotation occurs perpendicular to the boundary plane (Néel wall).

The coercivity can be enhanced by various methods:

- intentional addition of imperfections which impede the Bloch wall movement in large grains (e.g, carbides in steel magnets);
- preparation of single domain grains which can be switched only completely requiring high fields;
- implementation or enhancement of the magnet anisotropy (shape or crystal anisotropy).

Oriented AlNiCo 5 is an example of a magnet with shape anisotropy. When cooling down the molten components a spinodal decomposition into the magnetic phase FeCo and the less magnetic phase FeNiAl occurs. The spinodal decomposition as described by the Cahn–Hilliard equation is a diffusion process which results in a periodic and crystallographic oriented structure of the phases. The size of the segregated phases and the width of the boundaries (it is described by a tanh function) is a function of time.

When AlNiCo 5 is cooled down in the presence of a magnetic field the two phases get oriented and a strong shape anisotropy evolves [23]. As a consequence the energy product along the easy axis is about a factor of ten larger than in the perpendicular direction.

RE-based permanent magnets exhibit a large crystal anisotropy. These magnets are either of the nucleation type or the pinning type. Nucleation type magnets are $SmCo_5$, $Nd_2Fe_{14}B$, and ferrites. Pinning type magnets are $Sm(Co,Fe,Cu,Hf)_7$, $SmCo_5+Cu$ precipitation, and $Sm_2Co_{17}+SmCo_5$ precipitation of the size of domain wall thickness.

The Bloch walls of nucleation type magnets move easily within one grain and are stopped only at the grain boundaries. After heating, many domain walls exist within each grain. They are easily pushed out of the grain bulk to the boundaries with rather low fields (high initial permeability). Then, the domain walls are fixed to the grain boundaries. Once the magnet is fully magnetized, high reverse fields are needed to switch the domains. Most grains switch the magnetization in a single step without forming new Bloch walls.

Pinning type magnets have pinning centres within the grains which impede the Bloch wall motion. These can be impurities or precipitations. The two magnet types show a different initial magnetization after complete demagnetization by heating as plotted in Fig. 16.

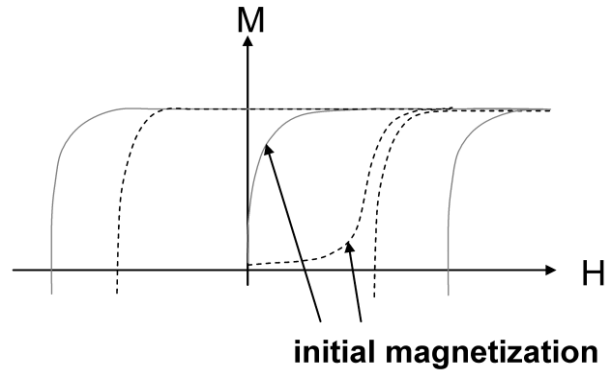

**Fig. 16:** Initial magnetization of nucleation type magnets (solid lines) and pinning type magnets (dashed lines)

Partial replacement of neodymium with dysprosium enhances the crystal anisotropy and, hence, the coercivity. Simultaneously, the remanence is reduced. $Nd_2Fe_{14}B$ magnet suppliers provide various grades of one material which differ in the Dy content. Depending on the specific application the appropriate material can be chosen. Dysprosium is rather expensive, so much effort has been spent in the maximization of the coercivity without using additional Dy. A correlation between the coercivity and the grain size was discussed by Mager in 1952 [24]. Systematic studies show a decrease of the coercivity with growing grain size within the range of the final grain size of 3.8 to 7.6 μm (Table 12):

**Table 12:** Coercivity dependence on grain size [25]

| Grain size before sintering (μm) | Final grain size (μm) | $H_{cj}$ (20°C) kA/m | $H_{cj}$ (100°C) kA/m |
|---|---|---|---|
| 1.9 | 3.8 | 1178 | 581 |
| 2.2 | 4.3 | 1162 | 573 |
| 2.6 | 4.9 | 1090 | 525 |
| 3.0 | 6.0 | 971 | 462 |
| 3.5 | 7.6 | 883 | 414 |

The experimental data can be fitted with $H_{cj}(20°C) \propto (\text{final\_grain\_size})^{-0.44}$. The grain size grows during sintering according to $R(t) = k \cdot t^{1/n}$.

The parameter $n$ depends on the material. For pure metals we find $n = 2–4$. In sintered $Nd_2Fe_{14}B$ the grain growth rate depends on the B concentration. It is about $n = 16–20$ for B concentrations below 5.7 at. % and decreases to $n = 7.5$ for B concentrations above 5.7 at. %. Also the RE-rich constituents have an impact on the grain growth rate. For magnets with RE-rich constituents < 4 wt. % we have $n = 30–40$. For higher fractions of RE-rich constituents $n$ decreases to $n = 10$ [26]. The sintering time has to be adjusted appropriately, to get an optimum grain size of 3–5 μm and to avoid giant grain growth. Usually, the grain size is measured according to the standard methods as defined in ASTEM E112.

Another figure of merit during sintering is the number of corners per grain (looking in the easy axis direction). Owing to the crystal structure of $Nd_2Fe_{14}B$ six corners indicate an unperturbed crystal structure. Hence the number of six-corner grains should be maximized [26].

The coercivity depends also on the alignment factor $f_\varphi$. With increasing $f_\varphi$ the remanence grows but the coercivity diminishes (Fig. 17) [20].

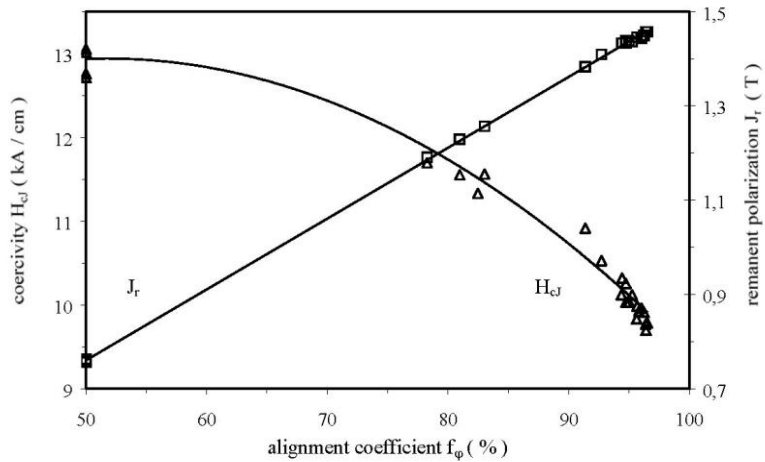

**Fig. 17:** Remanence and coercivity versus alignment factor [20]

The coercivity is a function of the direction of the applied external field [27]. It increases with the angle between the field and the easy axis. For angles smaller than 45° the dependence is roughly described by

$$H_{cj} \propto \frac{1}{\cos(\theta)}.$$

In certain cases this enhanced coercivity can be used in a magnetic design.

## 9 Observation of magnetic domains

An effective development of magnetic materials is based on detailed microscopic information of the material. A large variety of techniques is used for the observation of magnetic domains. In the following we can give only a brief overview of the most common methods and a few new methods. For more details we refer to the book by Schaefer (see bibliography).

Bitter elaborated a simple procedure to visualize magnetic domain boundaries by strewing ferromagnetic powder onto a magnetic surface. The ferromagnetic particles move to the areas of strong field gradients which are equivalent to the domain boundaries. The pictures are called Bitter patterns. Today, ferrofluids are used which are colloidal suspensions containing small ferromagnetic particles of the size of a few tens of nanometres. The resolution of this method is 100 nm. It is restricted to stationary measurements.

Fast processes can be studied by making use of various magneto-optical effects. All magneto-optical effects are described with a generalized permittivity tensor. For a cubic crystal it has the form

$$\vec{\varepsilon} = \varepsilon \begin{pmatrix} 1 & -iQ_V m_3 & iQ_v m_2 \\ iQ_v m_3 & 1 & -iQ_v m_1 \\ -iQ_v m_2 & iQ_v m_1 & 1 \end{pmatrix} + \begin{pmatrix} B_1 m_1^2 & B_2 m_1 m_2 & B_2 m_1 m_3 \\ B_2 m_1 m_2 & B_1 m_2^2 & B_2 m_2 m_3 \\ B_2 m_1 m_3 & B_2 m_2 m_3 & B_1 m_3^2 \end{pmatrix}$$

Similarly, a magnetic permeability tensor can be set up. The matrix elements, however, are two orders of magnitude smaller and are usually neglected. Inserting this tensor in the Fresnel equations, all of the magneto-optical effects can be described quantitatively. In magneto-optical spectroscopy a magnetic sample is irradiated with linearly polarized light. The rotation of the polarization vector in transmission geometry is called the Faraday effect and the rotation in reflection geometry is called the Kerr effect. The linear polarized light introduces vibrations of the charged particles in the sample. In the presence of a magnetic field the moving charges experience a Lorentz force. The modified vibration introduces perpendicular electric field components in the reflected or transmitted beam. Figure 18 shows the geometries of all magneto-optical effects.

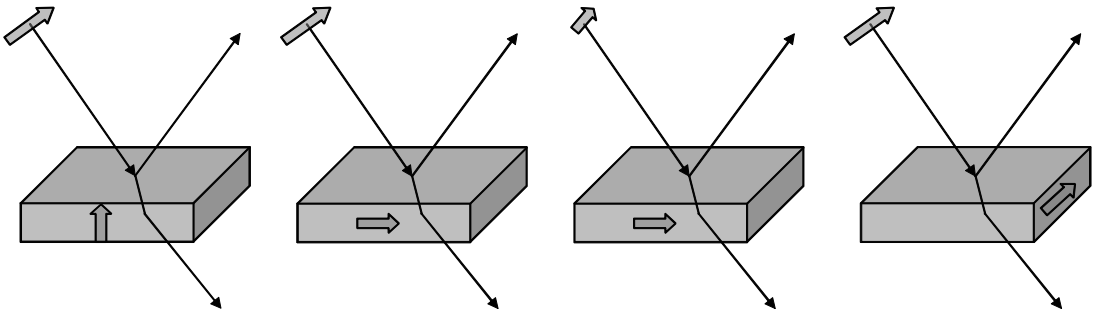

**Fig. 18:** Magneto-optical effects. From left to right: (i) Polar magnetization, in plane polarization: clockwise rotation of polarization in reflection and transmission. (ii) Longitudinal magnetization, in plane polarization: anticlockwise rotation of polarization in reflection and transmission. (iii) Longitudinal magnetization, polarization perpendicular to plane of incidence: clockwise rotation of polarization in reflection and anticlockwise rotation in transmission. (iv) Transverse magnetization, in plane polarization: in reflection no rotation but amplitude variation, no effect in transmission.

All these methods are not element specific. With the development of synchrotron radiation light sources the element specific investigation of magnetic samples in the soft X-ray regime has evolved to a large research field where the 2p–3d transitions of Fe, Co, Ni and the 3d–4f transitions of the rare earth metals are of particular interest. The samples are irradiated with circular polarized photons as produced by circular undulators in a storage ring. The different absorption coefficient of right and left circular polarized light is used to identify the magnetic domains (XMCD: X-ray Magnetic Circular Dichroism). Photoelectrons emitted by the sample can be used in a photoelectron microscope (PEEM) to visualize the local distribution of the domains (Fig. 19). It has to be emphasized that this type of experiment goes far beyond the usual Kerr effect measurement with visible light since the investigation of layered magnetic samples requires tuneable monochromatic light of variable polarization in the soft X-ray regime which is only available at modern synchrotron radiation light sources.

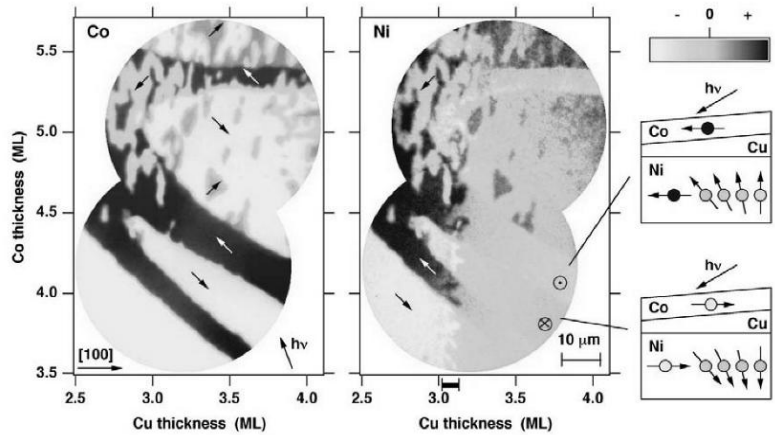

**Fig. 19:** Measurement of a system consisting of a Co layer on a Ni substrate separated by a wedged Cu layer making use of a photoelectron microscope at BESSY II. Here the exchange coupling between the Co and the Ni layer is studied in dependence of the thickness of the nonmagnetic Cu layer. At large distances the Ni atoms orient vertically, at small distances the Ni atoms tilt in the direction of the Co magnetic moments [28].

Magnetic domains can be imaged also with a soft X-ray holographic technique. A pinhole in the radiation cone of a circularly polarizing undulator prepares a coherent photon beam which illuminates the sample. Simultaneously, a small pinhole (100–350 nm) is irradiated with the same beam. Both transmitted beams are superimposed on a CCD camera. With the geometry of the reference pinhole known, the hologram can be inverted yielding the real structure of the magnetic domains of the sample [29]. No lenses, mirrors or zone plates are needed for this technique thus stressing the potential of this method. So far, a resolution of 50 nm has been demonstrated (Fig. 20).

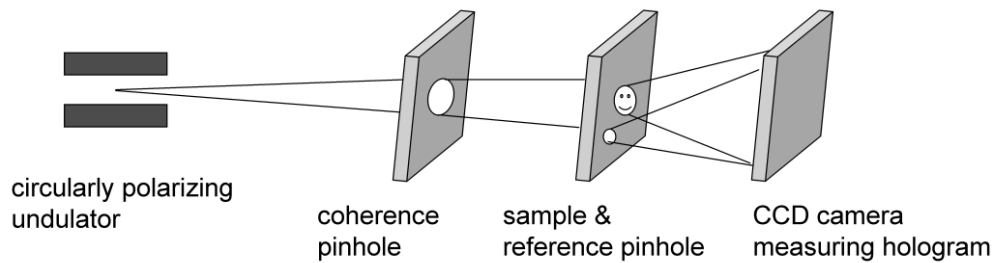

**Fig. 20:** Soft-X-ray holography using the circularly polarized undulator radiation. The magnetic sample is irradiated with right- and left-handed circular polarized light and the varying absorption strength depicts the orientation of the magnetization [29].

The bulk of thick magnetic samples can not be studied with photon beams on account of the limited penetration depth. Neutrons penetrate thick samples in the centimetre range. Neutron decoherence imaging is suited to detect magnetic domains of such samples [30]. The neutrons of a nuclear reactor hit a source grating. Each slit of the grating represents a coherent line source whereas the different line sources are mutually incoherent. A proper choice of the grating line density provides a coherent superposition of the fringes of all line sources in the image plane. After a few metres of free propagation the neutrons are scattered at the magnetic domains of a sample. A phase grating behind the sample imprints a periodic phase modulation onto the wavefront. During further propagation the phase modulation transforms into a density modulation (Talbot imaging) which can be detected with a sliding detection grating in front of a detector. An undistorted neutron beam produces a periodic intensity modulation. The modulation amplitude decreases for a distorted beam and gives information on the scattering process (Fig. 21). The resolution demonstrated so far is 50 – 100 µm.

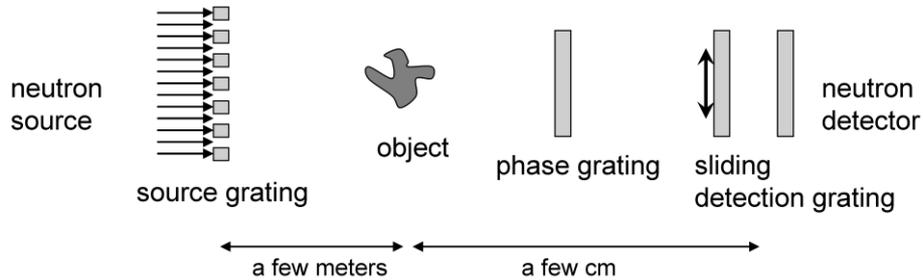

**Fig. 21:** Neutron decoherence imaging [30]

In plane magnetic domains thin layers can be detected with a transmission electron microscope acting as a Lorentz force microscope. Several 100 MeV electrons hit the magnetic film with magnetization vectors oriented perpendicular to the electron beam. The Lorentz force deflects the electrons and from their distribution behind the sample the magnetic domains can be reconstructed (Fig. 22). The resolution of this method is 10 nm. If the domains are oriented perpendicular to the film the sample has to be tilted to produce a deflection of the electrons. The resolution is reduced in these cases.

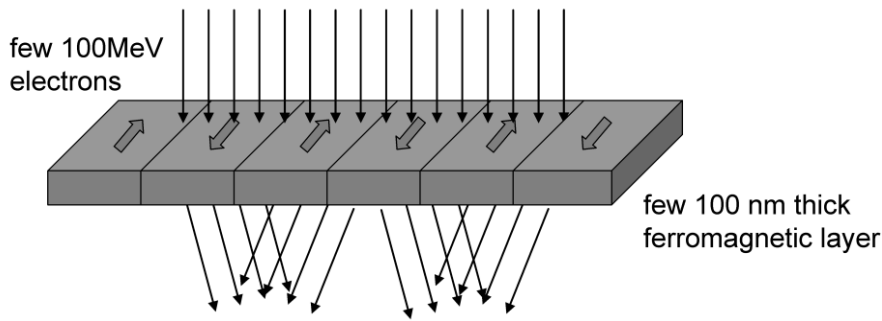

**Fig. 22:** Lorentz force microscope. For other domain geometries a tilting of the sample may be necessary

## 10  Simulation methods for permanent magnet devices

There are many codes on the market which solve Maxwell's equations in the presence of permanent magnets and current carrying wires, e.g., Refs. [31, 32]. In the following we concentrate on an algorithm which is very efficient for pure permanent magnet structures but also applicable to geometries including iron though it is significantly slower in these cases. The algorithm is widely used for the design of undulators and wigglers. The algorithm is implemented in the code RADIA [33, 34] which is freely available from the ESRF [35].

## 10.1 3D fields

Magnetic fields of pure permanent magnet structures can be simulated with an accuracy of a few per cent using the current sheet or charge sheet equivalent model (CSEM). These methods assume a permeability of one, hence the fields of individual blocks can be linearly superimposed. A magnet block is represented by current sheets at the magnet surfaces parallel to the easy axis or charge sheets at the surface perpendicular to the easy axis. The current sheets or charge sheets are assumed to be infinitely thin (Fig. 23).

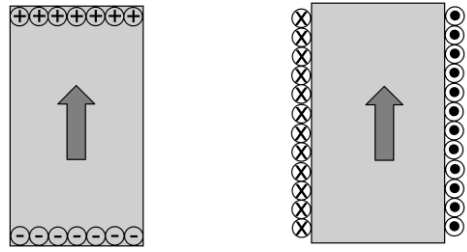

**Fig. 23:** Charge sheet (left) and current sheet (right) equivalent method

The magnetic induction of a current carrying wire is evaluated from the equation of Biot–Savart:

$$\vec{B}(\vec{r}_0) = \frac{1}{c} \int I d\vec{l} \times \frac{\vec{r}_0 - \vec{r}\,'}{|\vec{r}_0 - \vec{r}\,'|^3}.$$

In analogy in the current sheet equivalent method the magnetic induction is derived via integration over the surface current density $\vec{j}_M = \vec{M} \times \vec{n}$:

$$\vec{B}(\vec{r}_0) = \int \left(\vec{\nabla} \times \vec{M}(r')\right) \times \frac{\vec{r}_0 - \vec{r}\,'}{|\vec{r}_0 - \vec{r}\,'|^3} dV' = \iint_{surface} \frac{\vec{j}_M}{c} \times \frac{\vec{r}_0 - \vec{r}\,'}{|\vec{r}_0 - \vec{r}\,'|^3} dS'.$$

Following the charge sheet equivalent model a scalar potential is established by integrating over the surface charge density $\sigma_M = \vec{n}\cdot\vec{M}(\vec{r}\,')$ at the pole faces. The derivative of the scalar potential yields the magnetic field.

$$\Phi(\vec{r}_0) = -\int \frac{\vec{\nabla}'\cdot\vec{M}(r')}{|\vec{r}_0 - \vec{r}\,'|} dV' = \iint_{surface} \frac{\sigma_M \cdot dS'}{|\vec{r}_0 - \vec{r}\,'|}$$

$$\vec{H}(\vec{r}_0) = -\mathrm{grad}(\Phi(\vec{r}_0))$$

For rectangular blocks and a homogeneous magnetization the integration can be done analytically. In this specific case the magnetic induction is given by

$$\vec{B}(\vec{r}_0) = \overline{\overline{Q}}(\vec{r}_0) \cdot \vec{M}$$

$$Q_{xx} = \sum_{ijk=1}^{2} (-1)^{i+j+k+1} \arctan\left(\frac{y_j z_k}{x_i \sqrt{x_i^2 + y_j^2 + z_k^2}}\right)$$

$$Q_{xy} = \ln\left(\prod_{ijk=1}^{2} \left(z_k + \sqrt{x_i^2 + y_j^2 + z_k^2}\right)^{(-1)^{i+j+k}}\right)$$

$$x_{1,2}(y_{1,2}, z_{1,2}) = x_c(y_c, z_c) - x_0(y_0, z_0) \pm w_{x(y,z)}/2$$

and similar for the other $Q_{ij}$. The parameters $x_c$, $y_c$, $z_c$ define the centre of the magnet, $w_x$, $w_y$, $w_z$ are the magnet dimensions, and $x_0$, $y_0$, $z_0$ are the coordinates of the observation point.

Similarly, the magnetic induction $B$ and the field integral of any (plane) polygon can be evaluated [33, 34]. Based on these expressions the magnetic induction and field integrals of an arbitrary polyhedron can be evaluated:

$$\vec{B}(\vec{r}_0) = \overline{\overline{Q}}(\vec{r}_0) \cdot \vec{M}$$

$$\overline{\overline{Q}}(\vec{r}_0) = \oiint_{surface} \frac{(\vec{r}_0 - \vec{r}\,') \otimes \vec{n}\,'_{surface}}{\left|\vec{r}_0 - \vec{r}\,'\right|^3} d\vec{r}\,'$$

$$\vec{I}(r_0, \vec{v}) = \int_{-\infty}^{\infty} \vec{B}(\vec{r}_0 + \vec{v}) dl = \overline{\overline{G}}(\vec{r}_0, \vec{v}) \cdot \vec{M}$$

$$\overline{\overline{G}}(\vec{r}_0, \vec{v}) = 2 \cdot \oiint_{surface} \frac{[[(\vec{r}\,' - \vec{r}_0) \times \vec{v}] \times \vec{v}] \otimes \vec{n}\,'_{surface}}{\left|(\vec{r}\,' - \vec{r}_0) \times \vec{v}\right|^2} d\vec{r}\,'$$

$Q$ and $G$ are $3 \times 3$ matrices describing the geometric shape of the magnetic cell, $\otimes$ denotes a dyadic product and $\vec{v}$ is the integration direction. These equations are based on the assumption of a permeability of $\mu = 1$. Fields evaluated under this assumption are a few per cent higher than in reality. A higher accuracy is achieved with a realistic susceptibility $\chi \neq 0$ with $\mu = 1 + \chi$. The permeability of RE based permanent magnets is higher than one and it is different when parallel to or perpendicular to the easy axis. Typical values are $\mu_{par} = 1.06$, $\mu_{perp} = 1.17$. The values depend slightly on the fabrication procedure and the magnet grade. For example $\mu_{par} = 1.05$ for axially pressed and $\mu_{par} = 1.03$ for isostatic pressed magnets. $\mu_{par}$ shows no correlation with coercivity whereas $\mu_{perp}$ decreases with increasing coercivity (1.17 for $H_{cj} = 18$ kOe and 1.12 for $H_{cj} = 32$ kOe [27]).

To include a realistic permeability, iterative simulation strategies must be employed. The magnet configuration has to be segmented into individual cells where the cell size depends on the desired accuracy. Then, the geometry factors $Q$ and $G$ for the cells are evaluated. In a first run the magnetic induction and magnetic field $H_i$ at the centre of each cell is evaluated assuming a magnetization of $M_0 = M(H_i = 0)$ in each cell. In the following iterations the parallel and perpendicular magnetization is corrected and new values for the magnetic induction and magnetic field are derived:

$$\vec{B}_i = \sum_{\substack{k=1 \\ k \neq i}}^{N} \overline{\overline{Q}}_{k,i} \cdot \vec{M}_k + \overline{\overline{Q}}_{ii} \cdot \vec{M}_i$$

$$\vec{H}_i = \vec{B}_i - 4\pi \cdot \vec{M}_i$$

$$M_{i-par} = \frac{1}{4\pi} B_r + (\mu_{par} - 1) \cdot H_{i-par}$$

$$M_{i-perp} = (\mu_{perp} - 1) \cdot H_{i-perp}$$

A linear dependence of the magnetization on the magnetic field is assumed:

$$M_{par}(H_{par}) = M_r + \chi_{par} H_{par}$$

$$M_{perp}(H_{perp}) = \chi_{perp} H_{perp}$$

For reverse fields $H$ approaching $H_{cj}$ the magnetization does not change linearly anymore with $H$ and irreversible losses may occur. These cases can be simulated with a non-linear approach for the magnetization which includes also the temperature [36]. The temperature dependence of $M_r$, $H_{cj}$ and $\chi_{perp}$ can be parametrized as

$$M_r(T) = M_r(T_0) \cdot (1 + a_1(T - T_0) + a_2(T - T_0)^2 + ...)$$
$$H_{cj}(T) = H_{cj}(T_0) \cdot (1 + b_1(T - T_0) + b_2(T - T_0)^2 + ...)$$
$$\chi_{perp}(T) = \chi_{perp}(T_0) \cdot (1 + a_1(T - T_0) + a_2(T - T_0)^2 + ...)$$

where $a_i$ and $b_i$ are extracted from the data sheet of the magnet supplier. The non-linear behaviour of the magnetization for a given temperature $T$ is expressed by

$$M(H,T) = \alpha(T) \sum_{i=1}^{3} M_{si} \tanh\left(\frac{\chi_i}{M_{si}}(H - H_{cj}(T))\right) .$$

$M_{si}$, $\chi_i$ are derived from a fit of the $M(H)$ curve from the magnet supplier at $T = T_0$ and $\alpha(T)$ is determined from

$$M(H = 0, T) = M_r(T) .$$

This model has been tested at a real undulator structure and excellent agreement between simulation and measurement has been found [36].

It is worth noting that the operation of a permanent magnet in the third quadrant (still above the knee) does not imply that the magnet does not contribute to the field.

## 10.2 Complex notation of 2D field

The magnet geometry can be approximated with a two-dimensional model if the magnet is long compared to the end sections. Then, it is convenient to express the magnetic induction in complex notation:

$$\vec{B}^*(\vec{z}_0) = B_x - iB_y$$
$$\vec{z}_0 = x_0 + iy_0 = r_0 \cdot e^{i\varphi_0}$$

The complex conjugate $\vec{B}^*$ is used instead of $\vec{B}$ because $\vec{B}^*$ is an analytic function whereas $\vec{B}$ is not. Extremely useful tools for a magnet field optimization such as conformal mapping can be applied in this case (see next section). In complex notation the field of a current flowing into the plane is given by

$$\vec{B}^*(\vec{z}_0) = a \int \frac{j_z}{\vec{z}_0 - \vec{z}} \cdot dx \cdot dy .$$

Any iron-free permanent magnet distribution can be expressed as [37]:

$$\vec{B}^*(\vec{z}_0) = b \int \frac{\vec{B}_r}{(\vec{z}_0 - \vec{z})^2} \cdot dx \cdot dy ,$$
$$\vec{B}_r = B_{rx} + iB_{ry} .$$

This equation expresses an important rule: If the easy axis vector of all magnets of a complex two-dimensional configuration is rotated by the same angle $(+\varphi)$ this adds a factor of $e^{i\varphi}$ on the right-hand side. In consequence, the total field vector at a given point rotates by $(-\varphi)$: This behaviour is known as the **easy axis rotation theorem**.

Any source-free two-dimensional field distribution can be expanded in terms of multipoles. Using the complex notation we have a compact form with $a_n$ and $b_n$ being the regular and skew multipoles.

$$\vec{B}^*(\vec{z}) = \sum_{n=1}^{\infty}(a_n - ib_n)\vec{z}^{n-1}.$$

## 10.3 Conformal mapping

The technique of conformal mapping can be extremely useful in solving Maxwell's equations in complicated 2D-magnet arrangements [38–40]. Using a conformal map the complete configuration is transformed to a simpler geometry where Maxwell's equations are solved. Then, the results are transformed back to the old variables. One might argue that the technique of conformal mapping is no longer necessary when high performance multiprocessor computers are cheap and available everywhere. This is, however, only half the truth. For many problems the solution in transformed geometry provides higher accuracy for less effort, and, even more important, the conformal mapping technique provides a deeper insight into the properties of magnet systems. This understanding helps in a fast and efficient design.

Conformal maps preserve both angles and shapes of infinitesimal small figures, whereas the size of these figures usually changes. The intersection angle between any two curves in the original and the transformed geometry is identical.

Any analytic function represents a conformal map in the regions of non-zero derivatives. Let $F = \vec{w}(\vec{z})$ be a complex function with $\vec{w} = u(x, y) + iv(x, y)$ where $u$ and $v$ are real functions. $F$ is analytic if and only if the Cauchy–Riemann (C–R) relations are fulfilled:

$$\frac{\partial u}{\partial x} = \frac{\partial v}{\partial y}$$

$$\frac{\partial v}{\partial x} = -\frac{\partial u}{\partial y}$$

$u$ and $v$ also obey the Laplace equation which is proven by differentiating the C–R equations:

$$\frac{\partial^2 u}{\partial x^2} + \frac{\partial^2 u}{\partial y^2} = 0$$

$$\frac{\partial^2 v}{\partial x^2} + \frac{\partial^2 v}{\partial y^2} = 0$$

Within a source-free region (no currents, no magnetic charges) the complex magnetic field strength $\vec{H}^* = H_x - iH_y$ satisfies the C–R relations which are identical to Maxwell's equations in two dimensions. $\vec{H}^*$ is an analytic function and an integration yields the complex potential $F$:

$$F = A(x, y) + iV(x, y)$$

$$\vec{H}^*(\vec{z}) = i\frac{\partial F}{\partial \vec{z}}$$

$V(x,y)$ is the scalar potential and $A(x,y)$ the $z$ component of the vector potential in the usual sense, where $A_y$ and $A_x$ can be set to zero without losing generality (2-dimensional fields). As an example we will derive the conformal map that transforms a multipole into dipole geometry: For $F$ being the complex potential of $\vec{H}^*$ we have

$$H^*(\vec{z}) = i\frac{\partial F}{\partial \vec{z}} = i\frac{\partial F}{\partial \vec{w}}\frac{\partial \vec{w}}{\partial \vec{z}} = H^*(\vec{w}) \cdot \frac{\partial \vec{w}}{\partial \vec{z}} \ .$$

The field strength of an ideal regular octupole is given by $\vec{H}^*(\vec{z}) = o \cdot \vec{z}^3$ whereas a dipole is described by $\vec{H}^*(\vec{w}) = $ const. Thus we have $\vec{w}' = (o \cdot \vec{z}^3)/$ const and the conformal map that transforms the octupole to dipole geometry is given by $\vec{w}(\vec{z}) = (o \cdot \vec{z}^4)/(4 \cdot $ const$)$.

## 11    Permanent magnet multipoles

Accelerators are built of dipole magnets for the deflection of the relativistic particles, quadrupoles for a strong focusing of the electron beam, and sextupoles and higher order multipoles which cope with non-linear effects. Usually, the magnets are constructed as electromagnets since they provide flexibility in terms of the machine parameters. There are, however, cases where the flexibility is not needed and permanent magnets are the better choice. In the previous section we presented the fields of ideal multipoles. Now, we will discuss the layout of permanent magnet multipoles.

### 11.1  Fixed-strength multipoles

Halbach proposed segmented permanent magnet multipoles which approximate ideal multipoles to an accuracy limited by the number of magnets $M$ to be used per period [37]:

$$\vec{B}^*(\vec{z}) = \vec{B}_r \sum_{\nu=0}^{\infty} \left(\frac{\vec{z}}{r_1}\right)^{n-1} \frac{n}{n-1}\left(1 - \left(\frac{r_1}{r_2}\right)^{n-1}\right) K_n$$

$$K_n = \cos^n(\varepsilon\pi/M)\frac{\sin(n\varepsilon\pi/M)}{n\pi/M}$$

$$n = N + \nu M$$

$$\left.\frac{n}{n-1}\left(1-\left(\frac{r_1}{r_2}\right)^{n-1}\right)\right|_{n=1} = \ln(r_2/r_1)$$

Here, $N$ is the order of the multipole ($N = 1$: dipole, $N = 2$: quadrupole etc.), $\nu$ is the harmonic where $\nu = 0$ is the fundamental, $r_1$ and $r_2$ are the inner and outer radius, and $\varepsilon$ is the stacking factor. $A = (N + 1)2\pi/M$ is the relative angle of the easy axes of neighbouring magnet segments. A few examples are given in Fig. 24.

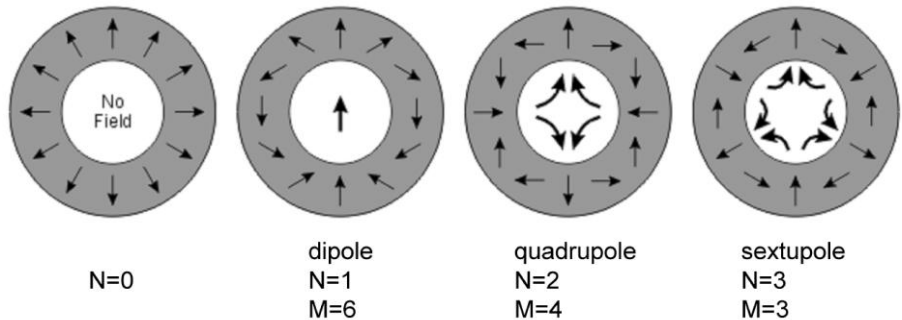

|   | dipole | quadrupole | sextupole |
| --- | --- | --- | --- |
| N=0 | N=1<br>M=6 | N=2<br>M=4 | N=3<br>M=3 |

**Fig. 24:** Halbach-type multipoles. From left to right: 1: $N = 0$, at the centre of the disk the field is zero according to the easy axis rotation theorem. 2: Dipole ($N = 1$, $M = 6$). 3: Quadrupole ($N = 2$, $M = 4$).  4: Sextupole ($N = 3$, $M = 3$).

Halbach-type permanent magnet quadrupoles ($M = 6$) with gradients of 500 T/mm in a 6 mm, 20 mm inner/outer diameter have been built [41, 42]. The higher order content of the quadrupoles could be significantly reduced by adjusting the individual magnet segments. Halbach multipoles can be improved with respect to the peak field when a fraction of the permanent magnets at the pole tips is replaced with soft iron (Fig. 25). A field gradient of 300 T/mm in a 14 mm inner diameter has been realized [43].

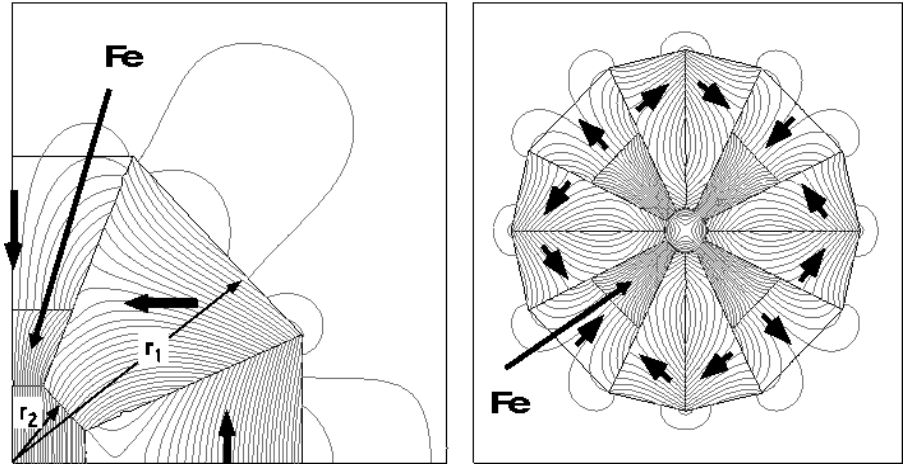

**Fig. 25:** Left: Modified Halbach dipole with iron pole tips. Right: Modified Halbach quadrupole (300 T/mm in the 14/100 mm inner/outer diameter) [43].

## 11.2 Variable strength multipoles

Permanent magnet quadrupoles are needed if the geometric constraints do not permit the use of bulky electromagnets such as at the final focus section of the ILC. The quadrupoles have to be placed close to the interaction point but must not block the other beam. A certain tuning range of the focusing strength is desirable which requires a specific design. Gluckstern proposed a continuously adjustable quadrupole consisting of five individual discs, each of them individually rotatable [44]. The linear effect of a rotated quadruple is given by a symplectic $4 \times 4$ matrix with

$$M^T \Phi M = \Phi = \Phi = \begin{pmatrix} 0 & 0 & 1 & 0 \\ 0 & 0 & 0 & 1 \\ -1 & 0 & 0 & 0 \\ 0 & -1 & 0 & 0 \end{pmatrix}$$

The off-diagonal elements characterize the coupling between the horizontal and vertical plane. With five individual discs with three rotation angles $\alpha$ and three quadruple strength $q$ (thicknesses) the coupling can be compensated and an arbitrary focusing can be realized: disc 1: $\alpha_1$, $q_1$; disc 2: $\alpha_2$, $q_2$; disc 3: $\alpha_3$, $q_3$; disc 4: $\alpha_2$, $q_2$; disc 5: $\alpha_1$, $q_1$ (Gluckstern singlet, Fig. 26, left). Prototypes with inner/outer diameters of 12/36 mm, respectively, and a gradient of 140 T/m have been built [45, 46].

Another design is the binary stepwise permanent magnet quadrupole [47] consisting of two layers of quadrupoles. The inner quadrupole is fixed in strength. The outer quadrupole is made of a series of rings with relative thicknesses of powers of two. Thus, a discrete (bitwise) adjustment is possible via rotating specific rings by multiples of $\pi/2$. The resolution is limited by the strength of the thinnest ring (Fig. 26, right).

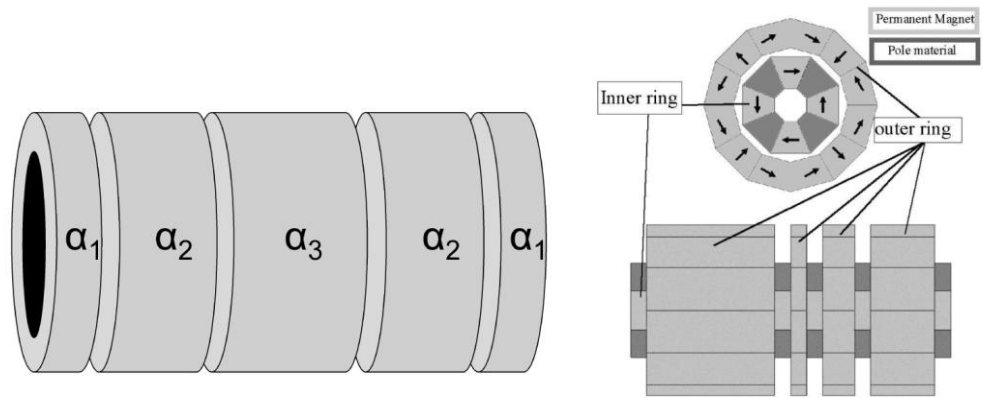

**Fig. 26:** Gluckstern singlet for continuous adjustment of quadrupole strength (left) [44] and binary stepwise adjustable quadrupole (right) [47]

## 12  Permanent magnet undulators and wigglers

Undulators and wigglers are period magnetic structures which force the electrons on a sinusoidal path. High field wigglers are used for electron beam damping to achieve low emittance in ultimate storage rings such as PETRA III [48] or NSLS II [49]. The main purpose of these devices, however, is the production of synchrotron radiation from the infrared to the hard X-ray region. First-generation synchrotron radiation light sources were storage rings which were shared between high-energy physics and synchrotron radiation users. Second-generation light sources were dedicated machines for the synchrotron radiation user community. Undulators and wigglers were still rarely used and bending magnets were the normal radiation sources. Third-generation synchrotron radiation light sources are explicitly built for the use of undulators and wigglers. The lattices have a high symmetry with several tens of straight sections to adapt as many undulators or wigglers as possible.

The brightness of photon beams as emitted by typical storage ring undulators is three to four orders of magnitude higher than that of dipole radiation (see Section 13). The photon energy is related to the undulator period length, and short-wavelength photons require short periods which can not be realized with electromagnets. Usually, permanent magnet undulators are chosen.

In 1973 Mallinson published a magnet design under the title: '*One-sided fluxes - A magnetic curiosity?*'. The article describes a linear array of magnets where the easy axis orientation is rotated between succeeding magnets. This array has a remarkable characteristic: At one side a significant amount of flux leaves the array whereas much less flux is detected at the opposite side. The high flux side is determined by the sense of rotation of the magnetization.

Halbach recognized the potential of this effect. He combined two of these arrays facing the high flux sides to each other. This array is called a pure permanent magnet undulator or Halbach I undulator [50] (Fig. 27).

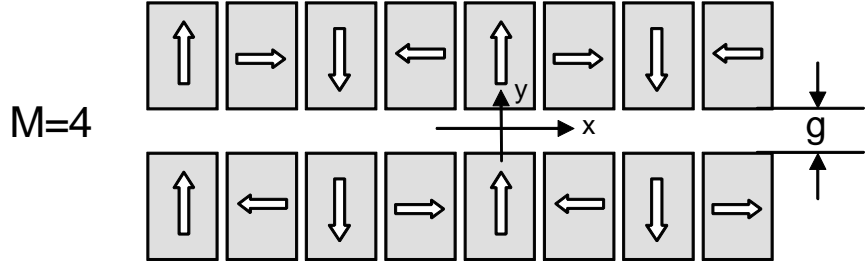

**Fig. 27:** Pure permanent magnet undulator (Halbach I)

The two-dimensional field of such an array can be expressed analytically:

$$\vec{B}^*(\vec{z}) = i2\vec{B}_r \sum_{v=0}^{\infty} \cos(nk\vec{z}) \cdot e^{-nkg/2} \cdot (1 - e^{-nkL}) \cdot \frac{\sin(n\varepsilon\pi/M)}{n\pi/M}$$

$$\vec{z} = x + i \cdot y$$

$$n = 1 + v \cdot M$$

$$k = 2\pi/\lambda_0$$

Here $\lambda_0$ is the period length, $\varepsilon$ is the filling factor, $g$ is the distance of the arrays (magnetic gap), $M$ is the number of magnets per period, and $n$ is the field harmonic. Note that the higher order field harmonic content is related to $M$. In most cases, four magnets per period are used. A larger $M$ enhances the field only by a few per cent. Higher fields can be produced with hybrid undulators, so-called Halbach II devices [51], where soft iron material is used to concentrate the flux (Fig. 28). For high performance applications the iron poles can be replaced by CoFe pieces with a saturation magnetization of 2.4 T to boost the peak magnetic field further.

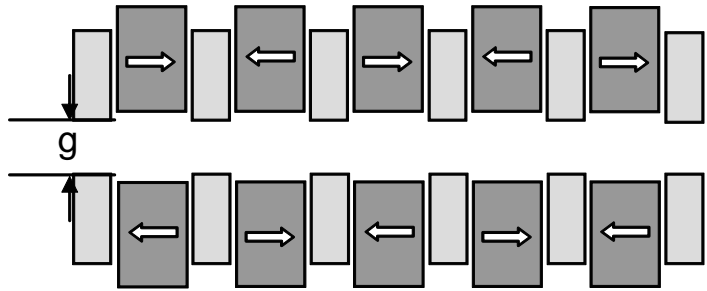

**Fig. 28:** Hybrid undulator (Halbach II)

There is no analytic field expression for a hybrid undulator and the field has to be parametrized based on numerical field simulations:

$$B_y \approx a \cdot \exp\left(-b \cdot \frac{g}{\lambda_0} + c \cdot \left(\frac{g}{\lambda_0}\right)^2\right).$$

For a hybrid undulator Elleaume derives the parameters $a = 3.69$, $b = 5.07$, $c = 1.52$ [52]. Parameters for other undulator designs are also presented in Ref. [52].

## 13 Spectral properties of undulators and wigglers

### 13.1 Ideal sources

The spectral properties of undulators and wigglers are described only briefly. For details we refer you to the books and CAS proceedings listed in the bibliography. The electromagnetic properties of relativistic particles are given by the Lienard–Wiechert potentials which are the solution of the inhomogeneous Maxwell equations (Gaussian units):

$$\Phi(\vec{x},t) = \left[\frac{e}{(1-\vec{\beta}\cdot\vec{n})R}\right]_{ret}$$

$$\vec{A}(\vec{x},t) = \left[\frac{e\vec{\beta}}{(1-\vec{\beta}\cdot\vec{n})R}\right]_{ret}$$

$\vec{\beta}(t_{ret})$ is the particle trajectory at the retarded time $t_{ret} = t - R(t_{ret})/c$, $\vec{n}(t_{ret})$ is the vector pointing from the particle to the observer, and $R(t_{ret})$ is the distance between the particle and the observer. The acceleration and velocity terms of the electric field are derived from these potentials and the pointing vector $\vec{S}$ describes the emitted power:

$$\vec{E}^{acc}(t) = \frac{e}{c}\cdot\left[\left(\vec{n}\times\left[(\vec{n}-\vec{\beta})\times\dot{\vec{\beta}}\right]\right)\bigg/\left(R\cdot(1-\vec{\beta}\cdot\vec{n})^3\right)\right]_{ret}$$

$$\vec{E}^{vel}(t) = e\cdot\left[(\vec{n}-\vec{\beta})\bigg/\left(\gamma^2\cdot R^2\cdot(1-\vec{\beta}\cdot\vec{n})^3\right)\right]_{ret}$$

$$\vec{B} = \left[\vec{n}\times\vec{E}\right]_{ret}$$

$$\vec{S} = \frac{c}{4\pi}\vec{E}\times\vec{B}$$

$\gamma$ is the Lorentz factor. A Fourier transform and a far field approximation deliver the spectral content of the radiation:

$$\frac{\partial^2 I}{\partial\omega\partial\Omega} = \frac{e^2}{4\pi^2 c}\left|\int_{-\infty}^{\infty}\left[\left(\vec{n}\times\left((\vec{n}-\vec{\beta})\times\dot{\vec{\beta}}\right)\right)\bigg/(1-\vec{\beta}\cdot\vec{n})^2\right]e^{i\omega(t-\vec{n}\cdot\vec{r})}dt\right|^2$$

In a vertical dipole field a charged particle describes a horizontal circle with a radius ρ. The radiation is given by the analytic expression

$$\frac{\partial^2 I}{\partial\omega\partial\Omega} = \frac{3e^2}{4\pi^2 c}y^2\gamma^2(1+(\theta\gamma)^2)^2\left[K_{2/3}(\xi)^2 + \left((\theta\gamma)^2\big/(1+(\theta\gamma)^2)\right)\cdot K_{1/3}(\xi)^2\right]$$

$$\xi = \frac{y}{2}(1+(\gamma\theta_y)^2)^{3/2}$$

$$y = \omega/\omega_c$$

$$\omega_c = (3\gamma^3 c)/(2\rho)$$

The first and the second part in rectangular brackets describe the contributions from the horizontal and vertical electric field component, respectively. In plane the radiation is horizontally polarized. Off plane a circular component shows up which is opposite in sign above and below the midplane.

Wigglers and undulators are periodic magnetic structures where the particles employ an oscillating trajectory. The devices are characterized by the undulator parameter $K = 93.4 \cdot \lambda_0 \cdot B_0$ where $\lambda_0$ and $B_0$ are the period length and the maximum field, respectively. Devices with $K \gg 1$ are called wigglers. The emitted radiation can be evaluated as a sum of dipole spectra emitted at each individual pole. In undulators the $K$ parameter and hence amplitudes of the oscillation are smaller and the radiation beams emitted at each of the poles overlap coherently. In planar undulators constructive interference is observed at the odd harmonics ($n = 1, 3, 5…$):

$$\lambda_n = \frac{\lambda_0}{n \cdot 2\gamma^2}\left(1 + K^2/2 + \gamma^2\theta^2\right)$$

Neglecting end pole effects, the undulator radiation in the far field close to the harmonics can be approximated with

$$\frac{\partial^2 I}{\partial\omega\partial\Omega} = \frac{e^2\gamma^2 N^2}{c} \cdot F_n(K_x, K_y, \gamma\theta, \gamma\Phi) \cdot \frac{\sin^2(N\pi \cdot \Delta\omega/\omega_1(\theta))}{N^2 \sin^2(\pi \cdot \Delta\omega/\omega_1(\theta))}$$

where $F$ is an infinite sum over Bessel functions.

## 13.2 Real sources

Wigglers have a low brightness and today they are used only for the production of high-energy photons which are not accessible with undulators. Wigglers are insensitive to field errors since the dipole spectra are spatially and spectrally broad and coherence effects can be ignored.

In contrast, undulators are based on a coherent overlap of the radiation contributions from the whole undulator. The spectral performance of an undulator is characterized by the phase error which describes the jitter in time between the electron beam and the emitted light:

$$\Delta\Phi = \frac{2\pi}{\beta\lambda(B\rho)^2} \cdot \int_0^z\left[\int_0^{z'} B_y^{fit} dz'' \cdot \int_0^{z'} B_y^{res} dz''\right] \cdot dz' + 0.5 \cdot \frac{2\pi}{\beta\lambda(B\rho)^2} \cdot \int_0^z\left[\int_0^{z'} B_y^{res} dz'' \cdot \int_0^{z'} B_y^{res} dz''\right] \cdot dz'.$$

Field errors produce phase errors which reduce the on-axis flux density of the odd harmonics approximately by $R = \left((1 - \exp(-\sigma_\phi^2))M + \exp(-\sigma_\phi^2)M^2\right)/M^2$ where $M$ is the number of poles [53]. Though the magnet quality of commercially available blocks has improved a lot during the last few years, it is still impossible to build high-quality permanent magnet undulators without applying specific techniques such as magnet sorting and undulator shimming (see Section 15). Phase errors below 2° can be achieved for devices with up to 100 periods. This limits the brightness reduction in the 15$^{th}$ harmonic to values well below 20%. Thus magnet field errors can be efficiently compensated and today the spectral performance of an undulator is only limited by the electron beam parameters, emittance, and energy spread.

## 14 Undulator and wiggler designs

There is a large variety of undulator and wiggler designs optimized for specific purposes. The device parameters can be grouped in four categories: tuning range, polarization, spectral purity, and on-axis-power density.

### 14.1 Tuning range: out-of-vacuum versus in-vacuum devices

In-vacuum undulators permit smaller period lengths than conventional devices for a fixed vertical vacuum aperture and $K$ parameter and, therefore, they provide higher photon energies. Planar in-vacuum devices rely on a mature technology and today many in-vacuum undulators are in operation all over the world [54].

As already mentioned, the negative temperature coefficients of the remanence and the coercivity of $Nd_2Fe_{14}B$ favour an operation at lower temperatures. Between 300 K and 150 K the remanence increases by 16%. The coercivity grows even more and thus another magnet grade with higher remanence and less coercivity can be used since the magnets gain stability at lower temperatures. Tanaka et al. proposed a cryogenic undulator to be operated at 150 K [55, 56]. Meanwhile, a cryogenic undulator is in operation at the ESRF [57–60], a 2 m device to be installed at the PSI [61] has been built at SPRING-8, and more devices are under consideration.

At 150 K $Nd_2Fe_{14}B$ features a spin reorientation [62, 63] and the remanence decreases below this temperature. By replacing the $Nd_2Fe_{14}B$ magnets with Pr-Fe-B magnets even lower temperatures can be used [55]. Temperatures below 80 K are of particular interest because textured dysprosium can be used as pole material which has an even higher saturation magnetization than CoFe. Furthermore, the temperature sensitivity of the magnetic performance decreases at lower temperatures. Recently, VAC in collaboration with Helmholtz-Zentrum Berlin (HZB) and Ludwig-Maximilian-Universitaet Muenchen (LMU) developed a new $(Pr,Nd)_2Fe_{14}B$ grade with an energy product of 65.3 MGOe at 85 K [64] (Fig. 29). A $(Pr,Nd)_2Fe_{14}B$ based undulator is currently under construction at HZB [65] and a short prototype Pr-Fe-B undulator has been measured recently in a vertical bath cryostat at the NSLS [66]. Tanaka et al. proposed to place passive high temperature superconducting loops around the pole tips of a cryogenic undulator. Closing the gap to 0 mm at a high temperature, cooling down and opening the gap again to the operation gap induces permanent currents in the loops which enhance the peak field further [67].

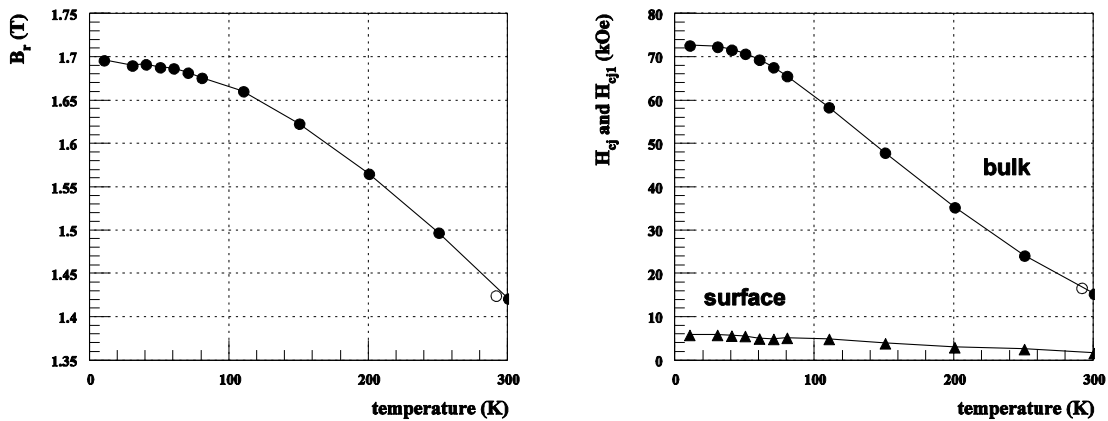

**Fig. 29:** Performance of $(Pr,Nd)_2Fe_{14}B$ material [64]

## 14.2 Variable polarization

The first wiggler for the production of linearly and elliptically polarized light, an asymmetric wiggler, was proposed in 1987 [68]. Positive and negative poles have different field amplitudes where the field integrals are compensated to zero. In the energy regime above the cut-off frequency of the weaker poles circularly polarized light is observed off plane just as in a bending magnet. The brightness of such sources is low due to the depth-of-field effect which increases with the angle of observation. However, high photon energies can be reached which are inaccessible with undulators. Asymmetric wigglers are being operated at several facilities [69–72]. At the ESRF an asymmetric wiggler with a peak field exceeding 3 T is in operation [73].

The brightness of an elliptical wiggler is significantly higher than that of an asymmetric wiggler since the on-axis beam is used [74]. Elliptical wigglers consist of individual magnet arrays for the horizontal and vertical magnetic fields where one of the field components is significantly larger than the other one. The electrons move on an elliptical trajectory producing elliptically polarized light on-axis. The helicity can be switched by moving one of the arrays longitudinally. Elliptical wigglers with variable helicity have been installed at the accumulator ring of the TRISTAN ring, at the Photon Factory [75, 76], and at SPRING-8 [77, 78].

In third-generation storage rings helical undulators are favoured on account of their higher degree of circular polarization and the higher brightness. Various helical undulator designs have been developed and implemented. The APPLE II design [79] provides the highest fields among all helical undulators and APPLE II undulators have become the work-horse in many light sources. The magnet arrays are split longitudinally and the individual magnet rows can be moved independently. In the elliptical mode (shifting two rows in the same direction) the horizontal and vertical fields are 90° out of phase whereas in the inclined mode (shifting two rows in opposite direction) they are in phase producing linearly polarized light at arbitrary angles. By moving three rows, any state of polarization can be produced and with an appropriate undulator setting polarizing effects of the beamline optical components in particular below 100 eV can be compensated [80, 81].

In the ESRF design [82] vertical and horizontal fields are produced by the upper and lower magnet girder, respectively. Any state of polarization can be realized and the good field region is large. However, the fields are lower than those of an APPLE II, the two beams have to be moved individually and the electron beam is steered vertically. The ELETTRA design [83] allows for a combined movement of both beams with one motor. This undulator provides only circularly polarized light (independent of gap setting), the fields are lower than in an APPLE II and the good field region is small. The SPRING-8 design [84] employs three magnet rows, the central one producing vertical fields and the side rows the horizontal fields. The good field region is large but the field levels are even lower than in the other designs.

## 14.3  Spectral purity: undulator periodicity

Usually, undulators have a periodic structure to achieve highest brightness at the fundamental and at higher harmonics. In the low-energy regime higher undulator harmonics may spoil the spectrum because they are efficiently transmitted through the monochromator together with the first harmonic. Helical devices produce only the first harmonic and higher harmonics are suppressed with the use of an appropriate pinhole. If the higher harmonics are needed and the spectral purity is essential, quasi-periodic structures as proposed by Sasaki et al. can help. The quasiperiodicity $z_m$ is derived from the projection of the grid points of a quadratic [85, 86] or rectangular lattice [87] with a grid parameter ratio of $r$ onto a tilted straight line with a tilt angle = $\alpha$:

$$Z_m = \frac{d}{r \cdot \tan(\alpha)} \left( m + (r \cdot \tan(\alpha) - 1) \cdot \left\lfloor \frac{\tan(\alpha)}{r + \tan(\alpha)} m + 1 \right\rfloor \right).$$

In the first design (Halbach I undulator with $M = 2$) the quasiperiodicity was realized with two different air gaps between the poles [88]. In a Halbach I $M = 4$ design the quasiperiodicity can also be introduced by completely removing longitudinally magnetized blocks, replacing these blocks with blocks of a reduced height [89, 90] or by retracting them [91]. Specific higher harmonics can be maximized with an appropriate choice of the amount of magnet retraction or height reduction. A successful suppression of higher harmonics has been demonstrated by measurements [92, 93]. A quasiperiodic hybrid undulator (Halbach II) has been installed at BESSY II [94]. The quasiperiodic scheme has also been applied to an electromagnetic undulator [95].

## 14.4  On-axis power density

All helical undulators have a reduced on-axis power density. This is a comfortable side effect, though these rather expensive devices are built for another purpose. A device emitting linearly polarized light with a reduced on-axis power density is the figure-eight undulator [96–99]. It consists of independent magnet arrays for the horizontal and vertical fields, respectively. One array has half the period length of the other one. In this way the projected electron trajectory has the shape of an eight. Such devices have been built at SPRING-8 [100–102] and at ELETTRA [103–105]. If an asymmetry is added to the figure-eight motion of the electrons the right- and left-handed circularly polarized light components no longer cancel and elliptically polarized light is emitted [106–108].

In 1990 Tatchyn [109] proposed a variable-period undulator because of the better performance as compared to a fixed-period undulator. If the photon wavelength could be tuned by changing the period length while keeping the undulator parameter $K$ constant ($K \approx 1$), highest brightness could be achieved without producing too much undesirable power in the higher harmonics. A variable-period undulator based on the staggered pole design [110] was proposed at the APS [111] but it was not built, finally, because of the mechanical complexity, even though the design has no permanent magnets. A variable-period ppm-undulator would be even more complicated. For an electromagnetic undulator a period doubling by rewiring has been demonstrated [112] and a superconducting undulator with the potential of period tripling has been proposed [113].

## 15 Shimming concepts for permanent magnet structures

Today's permanent magnet quality (dipole errors and magnetization homogeneity) is sufficient for most applications. There are, however, products which need a much better field performance and they require a sophisticated field tuning and optimization.

– Permanent magnet NMR spectrometers for medical applications require field homogeneities in the $10^{-6}$ regime which can be achieved only by shimming.

– A precise, smooth (low vibrations) and dynamic operation of a linear motor is essential for special applications. A homogeneous field quality can be achieved via magnet sorting.

– Sputtering facilities require a good field quality to provide homogeneous depositions. Magnet sorting and shimming help to reach the field accuracy.

– Permanent magnet based accelerator devices for third and next generation light sources such as multipoles or undulators need a sophisticated magnet sorting and shimming prior to installation.

In the following we concentrate on the shimming of undulators. The field optimization has to meet two targets: i) the spectral properties must not be deteriorated by magnet field errors, ii) the multipoles have to be sufficiently small to permit a transparent operation of the devices. The first goal can easily be achieved with well elaborated techniques whereas the second one needs more effort, in particular in the case of variably polarizing devices.

### 15.1 Static multipoles

Field errors of planar undulators are usually minimized (i.e., shimmed) with small soft iron sheets which are put on top of the magnets facing the electron beam. Trajectory errors are reduced with shims on transversely magnetized blocks whereas phase errors are minimized with shims on longitudinally magnetized blocks [114–117]. Another method developed by Pflueger varies the pole height for trajectory straightening [118].

Conventional shimming can not be applied in APPLE II undulators. Iron shims change their response when the magnet rows are phased and, furthermore, the air gap between the rows does not permit the application of conventional shims at the centre close to the electron beam. A combination of several techniques has to be applied to achieve a field quality comparable to planar devices. In the following we describe the technique as applied to the BESSY II devices [119].

Prior to assembly the three dipole components are measured in an automated Helmholtz coil system and the magnet block inhmogeneities are measured in a stretched wire set-up. Based on these data the magnets are sorted with a simulated annealing algorithm where the figure of merit is a combination of the transverse field integral distribution and the phase error. An excellent agreement between the prediction of the field integrals based on single block measurements and Hall probe measurements of the complete device is observed (Fig. 30).

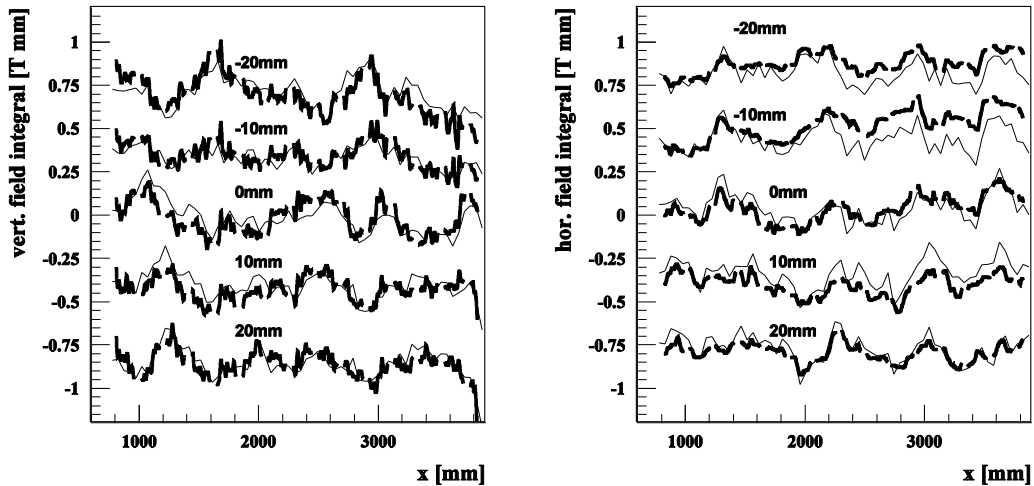

**Fig. 30:** Vertical (left) and horizontal (right) field integrals of the BESSY II UE49 APPLE undulator at various transverse positions. Thin lines: prediction from single block measurements. Thick lines: Hall probe scans of the complete device. The periodic part has been removed for a better visibility of the errors.

The pre-sorted and assembled structure is measured with a Hall probe, and trajectory and phase errors are removed via virtual shimming [120]. Here, the magnet blocks are moved transversely in the horizontal and vertical direction by up to 0.1 mm. Owing to the finite susceptibility of the magnetic material the field integrals change when phasing the magnet rows. This effect is minimized with iron shims which are placed on the vertically and longitudinally magnetized blocks. In the final step the phase independent field integrals are compensated with magic fingers at either end of the device. The magic fingers are arrays of small magnets with a cross-section of $4 \times 4$ mm$^2$ and variable thickness in the longitudinal direction. The thicknesses are derived from field integral measurements and the response functions of the magnets via a matrix inversion.

## 15.2 Dynamic multipoles

Undulators feature complicated three-dimensional magnetic fields which cause e-beam focusing (so-called edge focusing) and higher order dynamic effects which can not be described in terms of two-dimensional multipoles (the straight line integrals are zero). They do not obey two-dimensional Maxwell's equations and, sometimes, they are called pseudo-multipoles. The terms as derived in Ref. [121] have the form

$$\theta_{x/y} = -\frac{1}{(B\rho)^2}\int\left\{\int B_x dz' \cdot \int \frac{\partial B_x}{\partial x/y}dz' + \int B_y dz' \cdot \int \frac{\partial B_y}{\partial x/y}dz'\right\}dz$$

which for an undulator structure reduces to

$$\theta_{x/y} = -\frac{L}{2(B\rho)^2} \frac{\lambda_0^2}{(2\pi)^2} \left\{ B_x^0 \cdot \frac{\partial B_x^0}{\partial x/y} + B_y^0 \cdot \frac{\partial B_y^0}{\partial x/y} \right\}.$$

The pseudo-multipoles scale quadratically with the period length, the maximum field, and the inverse of the energy. Choosing too-small a pole width of a high field wiggler may result in beam dynamic effects which reduce the injection efficiency [122]. APPLE II undulators show strong horizontal field gradients, and hence high pseudo-multipoles. In the elliptical mode the terms can be compensated passively with L-shaped iron shims [123, 124]. In the inclined mode active compensation schemes have to be adopted. The low electron energy of BESSY II (1.7 GeV) requires an active compensation of the pseudo-multiples of the UE112 APPLE II undulator (112 mm period length). Twenty-eight flat wires are glued onto the vacuum chamber. They are powered individually and arbitrary transverse field integral distributions can be produced with a high accuracy [125, 126]. It is worth mentioning that the wires produce two-dimensional multipoles which, by principle, can not compensate the pseudo-multipoles. The wires or L-shims minimize the effects only in the midplane while adding non-linear terms out of plane. Here, they are less harmful if the vertical betatron function is smaller than the horizontal one.

## 16   Next-generation light sources

Today, many storage ring based third-generation light sources are in operation all over the world serving thousands of users every year. The fields of research include protein crystallography (aiming for an understanding of the function of large biological relevant molecules from its structure); investigation of magnetic materials employing magnetic circular dichroism (MCD) in the soft X-ray regime; element-specific microscopy on the nanometre scale; time-dependent spectroscopy down to the picosecond regime with low alpha optics; non-destructive 3D investigation of matter with respect to the chemical composition, morphology, internal stress and strain; high-pressure study of matter; investigation of electron correlation phenomena such as high-temperature superconductivity; archeometry; and many more topics.

In storage rings time-dependent studies can be extended down to the 100–200 fs using the femtosecond slicing technique [127]. A high-power infrared fs laser interacts with the electron bunch within an undulator (modulator) which is resonant to the laser. The energy of the particles in the interaction region is modulated. In a dispersive section the off-energy electrons are transversely displaced or deflected with respect to the unperturbed beam which permits a spatial separation of the two photon beams as produced in the next undulator (radiator). The polarization of the fs pulses is defined by the radiator. Femtosecond slicing facilities have been built at the ALS [128, 129], at BESSY II [130–132], and at the SLS [133]. At the ALS and the SLS the fs photon beam is linearly polarized whereas at BESSY II the polarization is variable (APPLE II).

The next-generation light sources are free electron lasers (FEL). They provide a high peak brightness, short pulses in the tens of fs regime, and longitudinal and transverse coherent light beams. These properties open new research areas. Currently, two soft X-ray FELs are in operation. FLASH at DESY provides photons up to 200 eV [134] and the SPRING-8 Compact SASE Source (SCSS) test accelerator [135] delivers photons up to 50 eV. FLASH uses fixed-gap undulators (the energy has to be tuned with the electron energy) whereas the SPRING-8 facility operates variable-gap in-vacuum undulators. In spring 2009 the first X-ray FEL, the Linac Coherent Light Source (LCLS), went into operation lasing at 0.15 nm [136]. The fixed-gap device has a transverse canting of the poles for a fine-tuning of the energy. Further X-ray FELs are under construction or are planned at DESY [137], at the SPRING-8 site [138], and at PSI [139]. These facilities will have several undulator beamlines with typical lengths of 100 m each. The total length of the first three European X-FEL undulator lines adds up to 555 m. The weight of the magnetic material of the three devices is 60 tonnes.

Another challenging linac-based light source concept is the energy recovery linac. A high-quality low-emittance beam as generated in an rf gun is injected into a circular machine and serves many users in parallel. Before the electron bunch is damped, which would be accompanied with growth of emittance and bunch length, it is recovered and replaced by a fresh bunch. ERLs provide a high averaged photon flux, a low emittance with a round beam, and short pulses in the 200 fs regime. Apart from the Jefferson Lab ERL [140], the existing and planned soft X-ray ERLs have a prototype character [141–146]. They are operated for the development and investigation of relevant technologies which are needed in X-ray ERLs. X-ray ERLs are planned at Cornell University [147], at the APS [148], and at KEK JAEA [149–150].

The new light sources require new undulator concepts. The period lengths of the undulators determine the electron energy and shorter period lengths permit shorter linacs, which reduces the total costs. Following this argument, in-vacuum undulators are planned at the SPRING-8 X-FEL and the PSI X-FEL. Circularly polarizing devices covering the regime up to 3 keV are under consideration. Studies on crossed undulators [151–153] predict a power which is one order of magnitude lower as compared to an APPLE II. The degree of polarization is only 80% whereas it is close to 100% for an APPLE II. Most challenging are variable polarizing devices operated under ultra high vacuum conditions. Temnykh built a 30 cm prototype of an in-vacuum Delta-magnet undulator [154]. The fixed-gap, variable polarization device has a four-fold symmetry without any access from the side. This is affordable in linac-based machines which do not require side access for injection. For similar gap-to-period length ratios the peak fields are comparable to the APPLE III design (Fig. 31) [155].

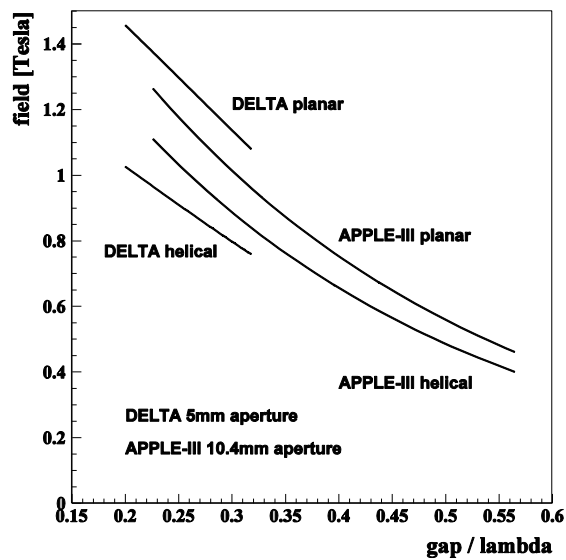

**Fig. 31:** Field comparison of a Delta undulator and an APPLE III

New acceleration concepts are used in so-called table-top FELs where electrons are accelerated up to the GeV regime within a length of a few centimetres [156]. A high-power laser ionizes the hydrogen atoms in a region of about 30 μm generating extremely high field gradients which accelerate the electrons. These devices are optimized for small dimensions, and short-period permanent magnet undulators as well as permanent magnet quadrupoles are essential. The small vertical apertures require magnets with high radiation stability. It is expected that in-vacuum undulators based on Pr-Fe-B material operated at low temperatures will play a key role in this field.

**References**

[1] Product leaflet of Vacuumschmelze 2007, http://www.Vacuumschmelze.com

[2] Y. Luo, 20th Workshop on Rare Earth Permanent Magnets, Crete, Greece, 2008.

[3] Y. Kaneko, 18th Workshop on HPMA, Annecy, 2004.

[4] Y. Iwashita et al., Proc. of the PAC, Portland, OR, USA (2003) 2198–2200.

[5] J. Osborn, Phys. Rev. 67 (1945) 351–357.

[6] A. Aharoni, J. Appl. Phys. 83 (1998) 3432–343.

[7] K. Bertsche et al., Proc of PAC, Dallas, TX, USA (1995) 1381–1383.

[8] P. F. Tavares et al., LNLS-2, Preliminary Conceptual Design Report, Campinas, Brazil, April 2009.

[9] W. Rodewald et al., Hagener Symposium für Pulvermetallurgie, Band 18 (2002) 225–245.

[10] Hitachi Metals Ltd., private communication, 2009.

[11] P. Colomp et al., Machine Technical Note 1-1996/ID, 1996.

[12] M. Petra et al., Nucl. Instrum. Methods. Phys. Res., A 507 (2003) 422–425.

[13] S. Sasaki et al., Proc. of the PAC, Knoxville, TN, USA, 2005, 4126–4128.

[14] H. Delsim-Hashemi et al., Proc. of the PAC, Vancouver, BC, Canada (2009).

[15] R. Qiu et al., Nucl. Instrum. Methods Phys. Res., A 594 (2008) 111.

[16] J. F. Herbst, Rev. Mod. Phys. 63 (1991) 819.

[17] T. Tanaka et al., Rev. Sci. Instrum. 71 (2000) 3010–3015.

[18] J. Chavanne, contribution to IMME 14, CERN, Geneva, Switzerland (2009).

[19] J. F. Herbst et al., Annu. Rev. Mater. Sci. 16 (1986) 467.

[20] W. Rodewald et al., IEEE Trans. Magn. 36 (2000) 3279–3281.

[21] M. Zakotnik et al., J. Iron Steel Res., International, 13 (2006) 289–295.

[22] M. Zakotnik et al., J. Alloys Comp. 450 (2008) L1–L3.

[23] J. Cahn, J. App. Phys. 34 (1963) 3581–3586.

[24] A. Mager, Ann. Phys (Leipzig), 446 (1952) 15–16.

[25] K. Üstüner et al., IEEE Trans. Magn. 42 (2006) 2897.

[26] W. Rodewald et al., IEEE, Trans. Magn. 39 (2003) 2932–2934.

[27] M. Katter, IEEE Trans. Magn. 41 (2005) 3853–3855.

[28] W. Kuch et al., Phys. Rev. B 65 (2002) 0064406.

[29] S. Eisebitt et al., Phys. Rev. B 68 (2003) 104419. S. Eisebitt et al., Nature 432 (2004) 885–888.

[30] F. Pfeiffer et al., Phys. Rev. Lett. 96 (2006) 215505. C. Grünzweig et al., Phys. Rev. Lett. 101 (2008) 025504.

[31] OPERA 2D and 3D: http://www.cobham.com/about-cobham/avionics-and-surveillance/about-us/technical-services.aspx

[32] MAXWELL: http://www.ansoft.com/products/em/maxwell/

[33] O. Chubar et al., J. Synchr. Rad. 5 (1998) 481–484.

[34] P. Elleaume et al., Proc. of the PAC, Vancouver, BC, Canada (1997) 3509–3511.

[35] RADIA: http://www.esrf.eu/Accelerators/Groups/InsertionDevices/Software

[36] J. Chavanne et al., Proc. of the EPAC, Vienna, Austria (2000) 2316–2318.

[37] K.Halbach, Nucl. Instrum. Methods, 169 (1980) 1–10.

[38] K. Halbach Nucl. Instrum. Methods, 64 (1968) 278.

[39] K. Halbach, Nucl. Instrum. Methods, 74 (1969) 147.

[40] K. Halbach, Understanding modern magnets through conformal mapping, Proc. Symposium on Conductivity and Magnetism: the Legacy of Felix Bloch, Stanford, CA, USA, 1989, Ed. W. A. Little (World Scientific, Singapore, 1990), pp. 1201–1222.

[41] T. Eichner et al. Phys. Rev. ST Accel. Beams, 10 (2007) 082401.

[42] S. Becker et al. Phys. Rev. ST Accel. Beams, 12 (2009) 102801.

[43] Y. Iwashita, Proc. of the PAC, Portland, OR, USA (2003) 2198–2200.

[44] R. Gluckstein et al., Nucl. Instrum. Methods, 187 (1981) 119–126.

[45] T. Sugimoto et al., Proc. of the EPAC, Genoa, Italy (2008) 583–585.

[46] Y. Iwashita et al., Proc of the PAC, Vancouver, BC, Canada (2009).

[47] Y. Iwashita et al., Proc. of the EPAC, Edinburgh, Scotland (2006) 2550–2552.

[48] PETRA III Technical Design Report, DESY 2004-035. http://petra3.desy.de/

[49] NSLS II Conceptual Design Report: http://www.bnl.gov/nsls2/project/CDR/

[50] K. Halbach, Nucl. Instrum. Methods, 187 (1981) 109–117.

[51] K. Halbach, J. Phys. Colloques, 44 (1983) C1-211–C1-216.

[52] P. Elleaume et al., Nucl. Instrum. Methods Phys. Res., A 455 (2000) 503–523.

[53] R. Walker, Nucl. Instrum. Methods Phys. Res., A 335 (1993) 328–337.

[54] T. Tanaka et al., Proc of the FEL Conference, Stanford, CA, USA (2005) 370–377.

[55] T. Hara et al., Phys. Rev. ST Accel. Beams, 7 (2004) 050702.

[56] T. Tanaka et al., J. Synchr. Rad. 14 (2007) 416–420.

[57] C. Kitegi et al., Proc. of the EPAC, Edinburgh, Scotland (2006) 3559–3561.

[58] J. Chavanne et al., Proc. of the EPAC, Genoa, Italy (2008) 2243–2245.

[59] J. Chavanne et al., Synchr. Rad. News 22 (2009) 34–37.

[60] J. Chavanne et al., Proc. of the SRI Conference, Melbourne, Australia (2009) to be published by the AIP.

[61] T. Tanaka et al., Phys. Rev. ST Accel. Beams, 12 (2009) 120702.

[62] J. Hu et al., Phys. Stat. Sol. b 188 (1995) 807.

[63] L. M. Garcia et al., Phys. Rev. Lett. 85 (2000) 429 and references therein.

[64] K. Üstüner et al., 20[th] Conference on Rare Earth Permanent Magnets, Crete (2008).

[65] J. Bahrdt et al., Proc. of the SRI Conference, Melbourne, Australia (2009), to be published by the AIP.

[66] T. Tanabe et al., Proc. of the SRI Conference, Melbourne, Australia (2009), to be published by the AIP.

[67] T. Tanaka et al., New J. Phys. 8 (1006) 287–303.

[68] J. Goulon, et al., Nucl. Instrum. Methods Phys. Res., A 254 (1987) 192–201.

[69] J. Pflueger et al., Nucl. Instrum. Methods Phys. Res., A 289 (1990) 300–306.

[70] J. Pflueger, Rev. Sci. Instrum. 63 (1992) 295–300.

[71] P. Elleaume et al., IEEE Trans. Magn. 28 (1992) 601.

[72] J. Chavanne et al., Rev. Sci. Instrum. 66 (1995) 1868–1871.

[73] J. Chavanne et al., Nucl. Instrum. Methods Phys. Res., A 421 (1999) 352–360.

[74] S. Yamamoto et al., Jpn. J. Appl. Phys. 26 (1987) L1613–L1615.

[75] S. Yamamoto et al., Rev. Sci. Instrum. 60 (1989) 1834–1837.

[76] S. Yamamoto et al., Phys. Rev. Lett. 62 (1989) 2672–2675.

[77] X. Marechal et al., Rev. Sci. Instrum. 66 (1995)1937–1039.

[78] X. Marechal et al., J. Synchr. Rad. 5 (1998) 431–433.

[79] S. Sasaki, Nucl. Instrum. Methods Phys. Res., A 347 (1994) 83–86.

[80] J. Bahrdt et al., Proc of the PAC, Vancouver, BC, Canada (2009), to be published.

[81] J. Bahrdt et al., Proc. of the SRI Conference, Melbourne, Australia (2009), to be published by the AIP.

[82] P. Elleaume, Nucl. Instrum. Methods Phys. Res., A 291 (1990) 371–377.

[83] B. Diviacco et al., Nucl. Instrum. Methods Phys. Res., A 292 (1990) 517–529.

[84] T. Hara et al., J. Synchr. Rad. 5 (1998) 426–427.

[85] S. Sasaki et al., Rev. Sci. Instrum. 66 (1995) 1953–1955.

[86] S. Hashimoto et al., Nucl. Instrum. Methods Phys. Res., A 361 (1995) 611–622.

[87] S. Sasaki, Technical Report ELETTRA, Trieste, Italy, ST/M–TN–98/24 (1998).

[88] M. Kawai et al., Proc. of the EPAC, Sitges (Barcelona), Spain (1996) 2549–2551.

[89] B. Diviacco et al., Proc. of the EPAC, Stockholm, Sweden (1998) 2216–2218.

[90] S. Sasaki et al., Proc. of the EPAC, Stockholm, Sweden (1998) 2237–2239.

[91] J. Chavanne et al., Proc. of the EPAC, Stockholm, Sweden (1998) 2213–2215.

[92] J. Chavanne et al., Proc. of the PAC, New York, NY, USA (1999) 2662–2664.

[93] B. Diviacco et al., Proc. of the EPAC, Vienna, Austria (2000) 2322–2324.

[94] J. Bahrdt et al., Nucl. Instrum. Methods Phys. Res., A 467–468 (2001) 130–133.

[95] T. Schmidt et al., Proc. of the EPAC, Paris, France (2002) 2631–2633.

[96] H. Kitamura, Rev. Sci. Instrum. 66 (1995) 2007–2010.

[97] T. Tanaka and H. Kitamura, Nucl. Instrum. Methods Phys. Res., A 364 (1995) 368–373.

[98] T. Tanaka and H. Kitamura, J. Synchr. Rad. 3 (1996) 47.

[99] T. Tanaka and H. Kitamura, J. Electron Spectrosc. Rel. Phenom. 80 (1996) 441–444.

[100]  T. Tanaka et al., J. Synchr. Rad. 5 (1998) 459–461.

[101]  T. Tanaka et al., Rev. Sci. Instrum. 70 (1999) 4153.

[102]  T. Tanaka et al., J. Appl. Phys. 88 (2000) 2101.

[103]  B. Diviacco et al., Proc. of the PAC, Chicago, IL, USA (2001) 2468–2470.

[104]  B. Diviacco et al., Proc. of the EPAC, Paris, France (2002) 2610–2612.

[105]  D. Zangrando et al., Proc. of the PAC, Portland, OR, USA (2003) 1050–1052.

[106]  T. Tanaka et al., Nucl. Instrum. Methods Phys. Res., A 449 (2000) 629–637.

[107]  T. Tanaka et al., Rev. Sci. Instrum. 73 (2002) 1724–1727.

[108]  K. Shirasawa et al., Phys. Rev. ST Accel. Beams, 7 (2004) 020702.

[109]  R. Tatchyn et al., IEEE Trans. Magn. 26 (1990) 3102–3123.

[110]  J. W. Lewellen et al., Nucl. Instrum. Methods Phys. Res., A 358 (1995) 24–76.

[111]  G. K. Shenoy, J. Synchr. Rad. 10 (2003) 205–213.

[112]  R. Klein et al., J. Synchr. Rad. 5 (1998) 451–452.

[113]  A. Bernhard et al., Proc. of the EPAC, Genoa, Italy (2008) 2231–2233.

[114]  D.C. Quimby et al., Nucl. Instrum. Methods Phys. Res., A 285 (1989) 281.

[115]  S.C. Gottschalk et al., Nucl. Instrum. Methods Phys. Res., A 296 (1990) 579.

[116]  B. Diviacco et al., Nucl. Instrum. Methods Phys. Res., A 368 (1996) 522.

[117]  P. Elleaume et al., Synchr. Rad. News 8 (1995) 18.

[118]  J. Pflüger et al., Nucl. Instrum. Methods Phys. Res., A429 (1999) 368.

[119]  J. Bahrdt et al., Nucl. Instrum. Methods Phys. Res., A 516 (2004) 575–585.

[120]  S. Marks et al., Proc. of the PAC, New York, NY, USA (1999) 162–164.

[121]  P. Elleaume, Proc. of the EPAC, Berlin, Germany (1992) 661–663.

[122]  J. Safranek et al., Proc. of the EPAC, Vienna, Austria (2000) 295–297.

[123]  J. Chavanne et al., Proc of the EPAC, Vienna, Austria (2000) 2346–2348.

[124]  J. Bahrdt et al., Proc. of the SRI Conference, Daegu, South Korea, AIP Conference Proceedings 879 (2006) 315–318.

[125]  J. Bahrdt et al., Proc of the EPAC, Genoa, Italy (2008) 2222–2224.

[126]  J. Bahrdt et al., Proc. of the PAC, Vancouver, BC, Canada (2009).

[127]  A. A. Zholents et al., Phys. Rev. Lett. 76 (1996) 912–915.

[128]  R. W. Schoenlein et al., Science, 287 (2000) 2237–2240.

[129]  R. W. Schoenlein, Appl. Phys. B 71 (2000) 1–10.

[130]  K. Holldack et al., Phys. Rev. ST Accel. Beams, 8 (2005) 040704.

[131]  K. Holldack et al., Phys. Rev. Lett. 96 (2006) 054801.

[132]  S. Khan et al., Phys. Rev. Lett. 97 (2006) 074801.

[133]  P. Beaud et al., Phys. Rev. Lett. 99 (2007) 174801.

[134] http://flash.desy.de/

[135] T. Tanaka et al., Proc. of the FEL Conference, Gyeongju, Korea (2008) 537–542. H. Tanaka et al., Proc. of the FEL Conference, Liverpool, UK (2009) 758–765.

[136] http://lcls.slac.stanford.edu/

[137] http://www.xfel.eu/de/

[138] http://www.riken.jp/XFEL/

[139] http://fel.web.psi.ch

[140] G. R. Neil et al., Nucl. Instrum. Methods Phys. Res., A 557 (2006) 9–15.

[141] L. Merminga, Proc. of the PAC, Albuquerque, NM, USA (2007) 22–26.

[142] E. Minehara et al., Nucl. Instrum. Methods Phys. Res., A 557 (2006) 16–22.

[143] V. P. Bolotin et al., Nucl. Instrum. Methods Phys. Res., A 557, (2006) 23–27.

[144] S. L. Smith, Proc. of the PAC, Albuquerque, NM, USA (2007) 1106–1108.

[145] V. Litvinenko et al., Proc. of the PAC, Albuquerque, NM, USA (2007) 1347–1349.

[146] M. Abo-Bakr, Proc. of the SRF Conference, Berlin, Germany (2009) 223–227.

[147] D. Bilderback et al., Synchr. Rad. News, 19, No. 6 (2006) 30–35. G.H. Hoffstaetter et al., Proc. of the PAC, Albuquerque, NM, USA (2007) 107–109.

[148] M. Borland et al., Proc. of the PAC, Albuquerque, NM, USA (2007) 1121–1123. N. Sereno et al., Proc. of the PAC, Albuquerque, NM, USA (2007) 1145–1147.

[149] T. Kasuga et al., Proc. of the PAC, Albuquerque, NM, USA (2007) 1016–1018.

[150] S. Sakanaka et al., Proc. of the EPAC, Genoa, Italy (2008) 205–207.

[151] Y. Ding et al., Phys. Rev. ST Accel. Beams, 11 (2008) 030702.

[152] Y. Li et al., Proc of the EPAC, Genoa, Italy (2008) 2282–2284.

[153] Y. Li et al., Nucl. Instrum. Methods Phys. Res., A (2009) accepted.

[154] A. Temnykh, Phys. Rev. ST Accel. Beams, 11 (2008) 120702.

[155] J. Bahrdt et al., Proc. of FEL Conference, Trieste, Italy (2004) 610–613.

[156] F. Grüner et al., Appl. Phys., B 86 (2007) 431–435.

**Bibliography**

R. J. Parker, Advances in Permanent Magnetism (Wiley, New York, 1990) ISBN 0–471–82293–0.

A. Hubert and R. Schäfer, Magnetic Domains: The Analysis of Magnetic Microstructures (Springer, Berlin, 1998).

H. Onuki and P. Elleaume, editors, Undulators, Wigglers and their Applications (Taylor and Francis, London, 2003) ISBN 0–415–28040–0.

J. A. Clarke, The Science and Technology of Undulators and Wigglers (Oxford University Press, New York, 2004) ISBN 0-19-850855-7.

R. Walker, Proceedings of CERN Accelerator School, Fifth Advanced Accelerator Physics Course, Rhodes, Greece, 1993, CERN 95-06 v2, editor S. Turner.

R. Walker, Proceedings of CERN Accelerator School, Synchrotron Radiation and Free Electron Lasers, Grenoble, France, 1996, CERN 98-04, editor S. Turner.

P. Elleaume, Proceedings of CERN Accelerator School, Synchrotron Radiation and Free Electron Lasers, Brunnen, Switzerland, 2003, CERN-2005-012, editor D. Brandt.

J. Bahrdt, Proceedings of CERN Accelerator School, Intermediate Accelerator Physics, DESY Zeuthen, Germany, 2003, CERN-2006-002, editor D. Brandt.